\documentclass[11pt,draftcls,onecolumn,twoside]{IEEEtran} 

\usepackage[latin1]{inputenc}
\usepackage[mathcal]{euscript}
\usepackage{amssymb,amsmath,amsthm}
\usepackage{graphicx}
\usepackage[english]{babel}
\usepackage{textcomp}
\usepackage{pstricks,pst-plot,pst-text,pst-tree,pst-eps,pst-fill,pst-node,
pst-math}

\title{Performance Analysis over Slow Fading Channels of a Half-Duplex
Single-Relay Protocol: Decode or Quantize and Forward}
\author{Nassar Ksairi$^{(1)}$\footnote{$^{(1)}$Sup\'elec, Plateau de Moulon
91192 Gif-sur-Yvette Cedex, France (nassar.ksairi@supelec.fr). 
Phone: +33 1 69 85 14 54, Fax: +33 1 69 85 14 69.}, 
Philippe Ciblat$^{(2)}$\footnote{$^{(2)}$CNRS / Telecom
ParisTech (ENST), 46 rue Barrault 75634 Paris Cedex 13, France
(ciblat@telecom-paristech.fr,bianchi@telecom-paristech.fr,walid.hachem@enst.fr).
Phone: +33 1 45 81 83 60, Fax: +33 1 45 81 71 44.}, 
Pascal Bianchi$^{(2)}$, Walid Hachem$^{(2)}$}

\newtheorem{theo}{Theorem}
\newtheorem{lemma}{Lemma}

\usepackage{color}
\definecolor{MyDarkBlue}{rgb}{0.1,0,0.65}
\definecolor{MyDarkRed}{rgb}{.85,0,0.1}

\begin{document}
\maketitle

\begin{abstract}
In this work, a new static relaying protocol is introduced for half duplex
single-relay networks, and its performance is studied in the context of
communications over slow fading wireless channels. The proposed protocol is
based on a \emph{Decode or Quantize and Forward} (DoQF) approach. In slow
fading scenarios, two performance metrics are relevant and complementary, 
namely the \emph{outage probability gain} and the \emph{Diversity-Multiplexing
Tradeoff} (DMT). 
First, we analyze the behavior of the outage probability $P_o$ associated with
the proposed protocol as the SNR $\rho$ tends to infinity. In this case, we
prove that $\rho^2 P_o$ converges to a constant $\xi$. We refer to this
constant as the outage gain and we derive its closed-form expression for a
general class of wireless channels that includes the Rayleigh and the Rice
channels as particular cases. We furthermore prove that the DoQF protocol has
the best achievable outage gain in the wide class of half-duplex static relaying
protocols. A method for minimizing $\xi$ with respect to the power distribution
between the source and the relay, and with respect to the durations of the slots
is also provided.

Next, we focus on Rayleigh distributed fading channels to derive the DMT
associated with the proposed DoQF protocol. Our results show that the DMT of
DoQF achieves the 2 by 1 MISO upper-bound for multiplexing gains $r<0.25$. 
\end{abstract}

\section{Introduction}
\label{sec:intro}
Relaying has become a widely accepted means of cooperation in wireless
communication networks. With this cooperation technique, the idle nodes that are
likely to be present in the vicinity of the transmitter can be used to relay
the source signal towards the destination, creating thus a virtual
Multiple-Input Multiple-Output (MIMO) system. In this paper, we focus on
networks composed of one source, one destination and one relay node that
operates under the half-duplex constraint \emph{i.e.,} the relay can either
receive or transmit, but not both at the same time. Under this assumption, the
relay listens to the source signal during a certain amount of time (the first
slot) and is allowed to transmit towards the destination during the rest of
the time (the second slot). We restrict our attention to \emph{static} relaying
protocols for which the relay listening time is fixed. This static property is
in contrast with \emph{dynamic} relaying protocols which allow the relay to
listen during a varying amount of time that depends on the (random) state of the
source-relay channel.

Recent works in relay-based cooperative wireless communications have proposed a
wide range of relaying protocols~\cite{NAF}-\cite{CF_Caire}. Most of these
protocols belong to one of the following families of relaying schemes: Amplify
and Forward (AF)~\cite{NAF,SAF}, Decode and Forward (DF)~\cite{NO_DF,DF,DDF} and
Compress and Forward (CF)~\cite{NO_DF,CF_DMT,CF_OFDM,CF_Caire}.
The first classical family of relaying protocols is formed by Amplify and
Forward (AF) protocols. In an AF setup, the relay retransmits a scaled version
of its received signal. Some of the most widespread amplify and forward
protocols are the Non orthogonal Amplify and Forward (NAF)~\cite{NAF} in case of
a single relay, and the Slotted Amplify and Forward (SAF)~\cite{SAF} in case of
multiple relays. By ``non orthogonal'' it is meant that the source and the relay
are simultaneously transmitting during the second slot.
A second well known family of protocols is formed by the Decode and Forward
(DF) approaches. In this case, the relay listens to the source during the first
slot of transmission and tries to decode the source message. If it succeeds, the
relay forwards the (re-coded) source message during the second slot. In this
context, Azarian~\emph{et al.}~\cite{DDF} proposed a dynamic version of the DF
(DDF, Dynamic Decode and Forward) in which the slots durations are supposed to
be adaptive as a function of the channel realization.  
Although the DDF is attractive from a theoretical point of view, an
implementation of the DDF requires the use of coders-decoders with adaptive
length. To the best of our knowledge, the design of such codes for the DDF is
still in its early stages~\cite{code_ddf1,code_ddf2,code_ddf3}. 
As stated in~\cite{code_ddf3}, the code designs proposed
in~\cite{code_ddf1,code_ddf2,code_ddf3} are not fully controllable in terms of
coding gain and entail very high decoding complexity when the frame length is
relatively large.
Recall that our focus in this paper is on static protocols \emph{i.e.,}
slot durations are assumed to be fixed w.r.t. channels realization. One of
the most widespread static DF protocols is the so-called \emph{non orthogonal}
DF~\cite{NO_DF} (as opposed to the \emph{orthogonal} DF~\cite{DF}). The non
orthogonal DF will be simply designated as DF in the rest of this paper. Hybrid
relaying strategies that can be considered as augmented DF protocols have also
been proposed. One example is the ``Amplify-Quantize-Decode-and-Forward
(AQDF)''~\cite{NO_DF}. In AQDF, a dedicated feedback link is assumed to exist
between the destination and the relay. Finally, another classical
relaying protocol is the Compress and Forward
(CF)~\cite{NO_DF,CF_DMT,CF_OFDM,CF_Caire}. In this protocol, the relay uses a
Wyner-Ziv encoder~\cite{Wyner_Ziv} to produce a source encoded version of its
received signal and forwards it assuming that the destination disposes of a side
information. This side information is the signal received on the direct
``source-destination'' link. It is worth mentioning that in the CF case, the
relay is assumed to have perfect knowledge of the channel gain between the relay
and the destination. Furthermore, some knowledge of the channel between the
source and the destination is also supposed available at the relay. Hybrid
strategies inspired by the CF scheme have also been proposed in the literature.
We cite for example~\cite{NO_DF} where the strong assumption of perfect
knowledge by the relay of the source-destination and the relay-destination
channels is replaced by a one-bit feedback link from the destination to the
relay. On the opposite, our work considers the context where both the channels 
``source to destination'' and ``relay to destination'' are completely unknown by
the relay, and where there are no feedback links in the network.

In this context, we propose a new relaying technique which we shall refer to as
the \emph{Decode or Quantize and Forward} (DoQF) protocol, and we analyze its
performance over slow fading wireless channels.
The DoQF can be considered as an augmented DF scheme, in which the
relay is able to adapt its forwarding strategy as a function of the information
that it received from the source during the first slot. 
More precisely, the relay first tries to decode the message of the source
based on the signal received during the first slot.
If the latter step is successful, then similarly to the classical DF scheme, the
relay retransmits a coded version of this message during the second slot based
on an independent codebook. In case the relay is not able
to decode the message, it does not remain inactive, but it quantizes the
received signal vector using a well chosen distortion value.

First, the DoQF has the advantage of a practical data
processing and receiver structure at both the relay and the destination.
Second, in the context of high-SNR transmission over slow fading channels, 
we demonstrate the optimality of the DoQF in a sense which is made clear below.

Assume that the source wants to transmit $R$ nats per channel use to
its destination, where constant $R$ is fixed w.r.t. the random channel
gains between the nodes of the network.  For a given value of the SNR
$\rho$, the outage probability $P_o(\rho)$ represents the probability that
the number of transmitted nats exceeds the mutual information
associated with the relay channel between the source and the
destination. Otherwise stated, $P_o(\rho)$ represents the probability
that the source message is lost. Generally speaking, the evaluation of
$P_o(\rho)$ for all possible values of the SNR $\rho$ is a difficult problem to
solve. For this reason, we focus on the high SNR regime. Indeed, as the SNR
$\rho$ tends to infinity, it is well known that informative expressions of the
outage probability can be derived. For instance, if the rate $R$ is a constant
w.r.t. the SNR $\rho$, it turns out that $\rho^2 P_o(\rho)$ usually converges to
a non-zero constant $\xi$ when $\rho$ tends to infinity. We will refer to this
constant $\xi$ as the \emph{outage gain}.
The outage gain provides crucial information about the behavior of the outage
probability in the high SNR regime. It is therefore a relevant performance
metric for the design of attractive relaying protocols.
In~\cite{outage_bound}, the authors optimize the power allocation for
an orthogonal DF protocol by minimizing an upper-bound on the outage
probability. In~\cite{outage_high_snr}, an AF protocol with one relay is
considered, and the power allocation is optimized by working on a high-SNR
approximation of the outage probability. Another approximation of the outage
probability at high SNR is considered by the authors
of~\cite{outage_high_snr_multi} to solve the problem of resource allocation for
an AF protocol with multiple relays. The factor $\xi$ associated with certain
relaying schemes was computed in a number of works in the literature (we
cite~\cite{coding_gain,coding_gain2} and~\cite{relay_power_alloc} without being
exclusive), but to the best of our knowledge, it has never been optimized  with
respect to the relaying protocols parameters. It is worth mentioning that the
protocols considered in all of the above contributions are orthogonal. Other
works propose methods to minimize the outage probability in the case where a
certain amount of instantaneous channel information is available through
feedback. This is the case for example
of~\cite{outage_feed1}~-~\cite{outage_feed3}.

Note that the derivation of the outage gain $\xi$ is based on the assumption
that the rate $R$ of the source is a constant w.r.t. the SNR $\rho$.
In practice, one could as well take benefit of an increasing SNR to increase
the transmission rate. When the rate $R=R(\rho)$ depends on the SNR,
a relevant performance metric in this case is the Diversity-Multiplexing
Tradeoff (DMT). The DMT was initially introduced by Zheng and Tse~\cite{tse} for
Rayleigh MIMO channels in order to capture the fundamental tradeoff between
diversity gain and multiplexing gain inherent to these channels. Since relay
channels can be considered as virtual MIMO systems, the same tool can be used as
a performance index for communications over Rayleigh distributed relay channels.
Following the definition of~\cite{DMG}, we shall write that a relaying protocol
achieves \emph{multiplexing gain} $r$ and \emph{diversity gain} $d(r)$ if the
rate $R(\rho)$ and the outage probability $P_o(\rho)$ associated with the
protocol satisfy: 
\begin{align}
\lim_{\rho\to\infty}\frac{R(\rho)}{\log\rho}=r& &\lim_{\rho\to\infty}\frac{\log
P_o(\rho)}{\log\rho}=-d(r)\:.\label{eq:DMT_def}
\end{align}
In this paper, $d(r)$ as defined above will be referred to as the DMT of the
relaying protocol. Note that the two performance metrics considered in this
paper, namely the outage gain and the DMT, are complimentary for the following
two reasons. First, the DMT is restricted to Rayleigh faded channels while the
outage gain has no such restriction. Second, we will see that the DMT of a
relaying protocol does not depend on the power partition between the source and
the relay, which is not the case of the outage gain.

The DMT has been used in the literature to evaluate the performance of different
relaying protocols over Rayleigh distributed fading channels. It is well known
that the DMT of any relaying scheme with a single relay is upper-bounded by the
DMT of a $2\times1$ MISO system which is given by 
$d_{\textrm{\footnotesize MISO}}(r)=2(1-r)^+$.
It has been shown in~\cite{DDF} that the DDF protocol achieves the MISO
upper-bound on the range of multiplexing gains $r<0.5$. As for the non
orthogonal DF, it is known from~\cite{DMG} that it does not achieve the MISO
bound for any multiplexing gain.
In the recent work~\cite{relay_dmt,inform_flow}, a new static protocol 
called ``quantize-map-and-forward'' is introduced and proved to achieve
the MISO upper-bound on the entire range of multiplexing gains. However, no
practical coding-decoding architecture has been proposed yet to implement this
recent protocol. Therefore, the design of DMT-optimal protocols which involve
practical transmit-receive architectures is still a challenging issue. In this
paper, we propose a protocol that has the advantage of both achieving the
MISO upper-bound on a part of the range of multiplexing gains and of being
implementable with practical coding-decoding architectures. Moreover,
simulations show that it has an excellent outage performance even for moderate
values of the SNR.\\

\noindent \emph{Contributions}

A novel DoQF relaying protocol for single-relay half-duplex networks is
introduced. The outage gain $\xi$ associated with the proposed DoQF
protocol is derived. A lower-bound on outage gains of the wide class of
half-duplex static protocols is also computed. The DoQF outage gain is shown to
coincide with the latter bound. Furthermore, a method to minimize $\xi$ with
respect to the protocol parameters is provided. Our simulations show that the
minimization of the outage gain is not only relevant in the high SNR regime, but
also over a wide range of SNR values, as it continues to reduce the outage
probability even for moderate values of $\rho$. The method proposed in this work
to derive $\xi$ does not make any assumption about the distribution of the
channels fading processes, except for the assumption that the probability
density of the channel gains does not vanish at zero. It can be shown that both
Rayleigh faded and Rice faded channels satisfy this assumption, and that only
the value at zero of the channel gains probability densities are needed by the
resource allocation unit. Finally, the closed-form expression of the DMT
associated with the DoQF protocol is provided. It is shown that the DoQF is
DMT-optimal for $r<0.25$ and outperforms the DMT of the DF protocol.\\

The rest of the paper is organized as follows.  A detailed description of
the new DoQF protocol is provided in Section~\ref{sec:description}.
The outage performance analysis and minimization at high SNR for a constant
transmission rate $R$ is addressed in Section~\ref{sec:outage}.
Theorem~\ref{the:DoQF_out} provides the closed-form expression of the outage
gain of the DoQF protocol. The minimization of this outage gain with respect to
the protocol parameters is next addressed in Subsection~\ref{sec:opt}. 
Section~\ref{sec:DoQF_DMT} is devoted to the DMT analysis of the DoQF protocol.
The main result of this section is presented in Theorem~\ref{the:DoQF_DMT} which
gives the closed-form expression of the DMT of the DoQF. Numerical results of
the outage gain and the DMT of the proposed protocol are drawn in
Section~\ref{sec:sim}. Finally, Section~\ref{sec:concl} is devoted to the
conclusions.\\

\noindent \emph{General Notations and Assumptions}

Before going further, we give the general notations and channel assumptions used
throughout the paper. In the sequel, node 0 will coincide with the source, node
1 with the relay and node 2 is the destination. The wireless channels between
the different nodes of the network are assumed to be independent channels and we
denote by $H_{ij}$ the complex random variable representing the wireless channel
between node $i$ and node $j$ with $i,j\in\{0,1,2\}$ (in this paper, scalar and
vector random variables are represented by upper case letters). Channel
coefficients $H_{ij}$ are assumed to be perfectly known at the receiving node
$j$, but are unknown at each other node of the network, including the
transmitter $i$. The power gain of the channel between node $i$ and node $j$
will be denoted by $G_{ij}=|H_{ij}|^2$. Notation $\mathcal{CN}(a,\sigma^2)$
stands throughout the paper for the complex circular Gaussian distribution with
mean $a$ and variance $\sigma^2$ per complex dimension.

Given two events $\mathcal{E}_1$ and $\mathcal{E}_2$, \emph{i.e.,} two
measurable subsets of a probability space $\Omega$, we denote by
$\mathrm{Pr}[\mathcal{E}_1]$ the probability measure of $\mathcal{E}_1$ and by
$\mathrm{Pr}[\mathcal{E}_1, \mathcal{E}_2]$ the probability of the intersection
of $\mathcal{E}_1$ and $\mathcal{E}_2$. We also write as usual
$f(\rho)\stackrel{.}{=}\rho^d$ if 
$\lim_{\rho\to\infty}\frac{\log f(\rho)}{\log(\rho)}=d$.
Notations $\stackrel{.}{>}$, $\stackrel{.}{<}$ are similarly defined. Finally,
$(x)^+=\max(0,x)$.

\section{The Proposed DoQF Protocol}\label{sec:description}
\label{sec:DoQF_descrip}
\subsection{Description of the Protocol}

The source (node~0) needs to send information at a rate of $R$ nats per channel
use towards the destination (node~2).
The source has at its disposal a frame of length $T$ and a dictionary of
$\left\lfloor e^{RT}\right\rfloor$ Gaussian independent vectors with independent
$\mathcal{CN}(0,1)$ elements each. We partition the word $X_0$ selected by the
source as $X_0=\left[X_{00}^\mathrm{T},X_{01}^\mathrm{T}\right]^\mathrm{T}$
where the length of $X_{00}$ and $X_{01}$ is $t_0 T$ and $t_1 T$ respectively
with $t_1=1-t_0$. Here $t_0<1$ is a fixed parameter. The source transmits the
vector $\sqrt{\alpha_{0}\rho} X_0=\left[\sqrt{\alpha_{0}\rho} X_{00}^\mathrm{T},
\sqrt{\alpha_{0}\rho} X_{01}^\mathrm{T}\right]^\mathrm{T}$,
where $\rho T$ represents the total energy spent by both the source and the
relay (node~1) to transmit the message as will be clear later on. The factor
$\alpha_0$ is the amplitude gain applied by the source, which means that 
$E_0=\alpha_0 \rho T$ is the source share of the total energy available for the
transmission of the block of $RT$ nats. Denote by $E_1$ the \emph{average}
energy spent by the relay for the transmission. The energy $E_1$ should be
selected such that the following constraint is
respected
\begin{equation}\label{eq:contrainte}
E_0+E_1 \leq \rho T\:.
\end{equation}
The selection of $E_1$ which does not violate the above constraint will be
addressed in Sections~\ref{sec:outage} and~\ref{sec:DoQF_DMT}.
Note that due to the fact that $E_1$ is defined as an average energy, 
constraint~\eqref{eq:contrainte} is equivalent to a long-term power constraint.

\begin{figure}[h]
\centering\includegraphics[width=8cm]{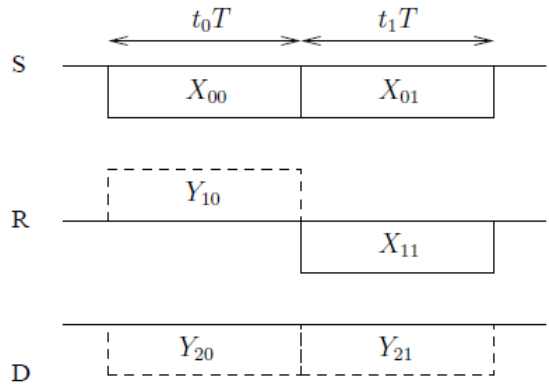}
\caption{Transmit/Receive signals for source (S), relay (R) and destination (D)}
\label{fig:slots}
\end{figure}

The relay listens to the source message for a duration of $t_0 T$
channel uses. At the end of this period of time that we refer to as slot~0, the
signal of size $t_0 T$ received by the relay writes
\begin{equation}\label{eq:y10}
Y_{10}=\sqrt{\alpha_{0}\rho}H_{01}X_{00}+V_{10}\:,
\end{equation}  
where each component of vector $V_{10}$ is a unit variance Additive White
Gaussian Noise (AWGN) at the relay. Figure~\ref{fig:slots} represents the
transmit and receive signals respectively for each node of the network. 
We now consider separately the case when the relay manages to decode the source
message and the case when it does not. 

\noindent {$\bullet$ \bf Case when the relay decodes the source message}

By referring to~\eqref{eq:y10}, we can check that the relay is able to decode
the source message if the event 
\begin{equation}\label{eq:event_E}
\mathcal{E}=\{\omega:t_0\log(1+\alpha_0\rho G_{01}(\omega))>R\}
\end{equation} 
is realized. If this is the case, the relay transmits during the remainder of
the frame (slot~1) the corresponding codeword of length $t_1 T$ from its own
codebook. The relay codebook is assumed to be independent from the codebook of
the source and is composed of $\left\lfloor e^{RT}\right\rfloor$ Gaussian
independent vectors with independent $\mathcal{CN}(0,1)$ elements each.
We denote by $X_{11}$ the codeword selected by the relay. The latter transmits
thus the vector $\sqrt{\alpha_1\rho}X_{11}$, where the factor $\alpha_1$ is the
amplitude gain applied by the relay. This means that $\alpha_1 \rho T$ is the
relay share of the total energy available for the transmission. Finally, during the 
slots 0 and 1, the destination
receives the signal
\begin{equation}
[Y_{20}^\mathrm{T},Y_{21}^\mathrm{T}]^\mathrm{T}=
{\bf H}_{\mathcal{E}}[X_{00}^\mathrm{T},X_{01}^\mathrm{T},X_{11}^\mathrm{T}]
^\mathrm{T}+[V_{20}^\mathrm{T},V_{21}^\mathrm{T}]^\mathrm{T}\:,
\end{equation}
where 
$${\bf H}_{\mathcal{E}}=\left[\begin{array}{lll}
\sqrt{\alpha_0\rho}H_{02}{\bf I}_{t_0 T}& 0& 0\\
0& \sqrt{\alpha_0\rho}H_{02}{\bf I}_{t_1 T}& \sqrt{\alpha_1\rho}H_{12}{\bf
I}_{t_1 T}
\end{array}\right]\:,$$ 
and where components of vector $V_{20}$ (resp. $V_{21}$) are unit variance AWGN
at the destination during slot~0 (resp. slot~1). 


\noindent {$\bullet$ \bf Case when the relay does not decode the source
message}

This is the case when the event $\overline{\mathcal{E}}$ is realized.
The relay quantizes in this case the received signal during slot~0 and transmits
a coded version of the quantized vector towards the destination during slot~1
using the following steps.\\
\emph{a) Quantization}: Denote by $\tilde{Y}_{10}$ the quantized version
of the received vector $Y_{10}$. Vector $\tilde{Y}_{10}$ is constructed
as follows. Clearly, all $t_0 T$ components of vector $Y_{10}$ are
independent and $\mathcal{CN}(0,\alpha_0\rho G_{01}+1)$ distributed. Denote
by $\Delta^2(\rho)$ the desired squared-error distortion per vector component:
$$\mathbb{E}|\tilde{Y}_{10}(i)-Y_{10}(i)|^2\leq\Delta^2(\rho)\:.$$ 
It is clear that letting the quantization squared-error depend on the SNR $\rho$
provides us with an additional degree of freedom in the design of the protocol
as we will see. The Rate Distortion Theorem for Gaussian sources~\cite{cover}
tells us that there exists a $(\left\lfloor e^{Q(\rho) t_0 T}\right\rfloor,t_0
T)$-rate distortion code (for some $Q(\rho)>0$) which is achievable for the
distortion $\Delta^2(\rho)$ if the following condition is satisfied
\begin{equation}\label{eq:Q_cond}
Q(\rho)>\log\left(\frac{\alpha_0\rho G_{01}+1}{\Delta^2(\rho)}\right)\:.
\end{equation}
In practice, such a code can be constructed by properly selecting the quantized
vector $\tilde{Y}_{10}$ among a quantizer-codebook formed by $\left\lfloor
e^{Q(\rho) t_0 T}\right\rfloor$ independent random vectors with distribution 
$\mathcal{CN}(0,(\alpha_0\rho G_{01}+$ 
$1-\Delta^2(\rho))\mathbf{I}_{t_0 T})$. 
Vector $\tilde{Y}_{10}$ is selected from this codebook in such a way that
sequences $Y_{10}$ and $\tilde{Y}_{10}$ are jointly typical w.r.t. the joint
distribution $p_{(Y,\tilde{Y})}$ given by 
\begin{equation}\label{eq:y_y}
Y=\tilde{Y}+\Delta(\rho) Z\:,
\end{equation}
where $\tilde{Y}$ and $Z$ are independent random variables with respective
distributions $\mathcal{CN}(0,\alpha_0\rho G_{01}+1-$ $\Delta^2(\rho))$
and $\mathcal{CN}(0,1)$. Condition~\eqref{eq:Q_cond} ensures that such a vector
$\tilde{Y}_{10}$ exists with high probability as $T\to\infty$.

Parameter $Q(\rho)$ can be interpreted as the number of nats used to quantize
one component of the received vector $Y_{10}$. This parameter must be chosen
such that condition~\eqref{eq:Q_cond} is satisfied. As the rhs
of~\eqref{eq:Q_cond}  depends on the channel gain $G_{01}$, it looks
impossible at first glance to construct a fixed quantizer which is successful
for any channel state. Nevertheless, recall that we are considering the case
where event $\mathcal{E}$ is \emph{not} realized \emph{i.e.,} $t_0
\log(1+\alpha_0\rho G_{01})<R$. In order to guarantee that
condition~\eqref{eq:Q_cond} always hold, it is thus sufficient to define
\begin{equation}\label{eq:Q}
Q(\rho)=\log\left(\frac{K}{\Delta^2(\rho)}\right)\:,
\end{equation} 
where $K$ is any constant such that $K\geq e^{\frac{R}{t_0}}$. We choose
$K=e^{\frac{R}{t_0}}$. 
In order to complete the definition of the quantizer, we still need to define
the way $\Delta^2(\rho)$ depends on the SNR $\rho$. This issue is
addressed at the end of the current section.\\
{\bf Remark:} Note that condition~\eqref{eq:Q_cond} implies that the following
inequality should hold
\begin{equation}\label{eq:Q_cond_b}
\alpha_0\rho G_{01}+1>\Delta^2(\rho)\:.
\end{equation}
Condition~\eqref{eq:Q_cond_b} is indeed necessary for the construction of the
quantization code because it ensures that the variance 
$\alpha_0\rho G_{01}+1-\Delta^2(\rho)$ of each component of the codewords is
positive. The quantization step is thus possible provided that the event 
\begin{equation}\label{eq:event_S}
\mathcal{S}=\left\{\omega:\alpha_0\rho G_{01}(\omega)+1>\Delta^2(\rho)\right\}.
\end{equation} 
is realized. In case the complementary event $\overline{\mathcal{S}}$ is
realized, the relay does not quantize the source message and remains silent
during slot~1. The latter case happens with negligible probability provided that
$\Delta^2(\rho)$ is chosen properly.\\
\emph{b) Forwarding the Relay Message}: During the second slot of length 
$t_1 T$, the relay must forward the index of the quantized vector among the
possible $\left\lfloor e^{Q(\rho)t_0 T}\right\rfloor$ ones. To that end, it
uses a Gaussian codebook with rate $Q(\rho) t_0/t_1$. If we denote by $X_{11}$
the corresponding codeword, the signal transmitted by the relay can be written
as $\sqrt{\phi(\rho)}X_{11}$, where $\phi(\rho)$ is the power of the relay.

Function~$\phi(\rho)$ should be selected such that the power constraint given
by~\eqref{eq:contrainte} is respected. A possible choice would be 
$\phi(\rho)=\alpha_1\rho$, which is the same power that
the relay spends when event $\cal E$ is realized. In this case, the relay
transmits during slot~1 at the same constant power $\alpha_1\rho$ regardless of
the fact that the source message has been decoded or not. Of course, the
factor $\alpha_1$ should be fixed in this case such that
constraint~\eqref{eq:contrainte} is respected. While this choice for
$\phi(\rho)$ is relatively simple, other possible choices which may lead to
better performance of the DoQF without violating the average power constraint
are discussed at the end of the current section.\\ 
\emph{c) Processing at Destination}: 
In case the relay has quantized the source message (event $\mathcal{S}$ defined
by~\eqref{eq:event_S} is realized), the destination proceeds as follows. It
first tries to recover the relay message $X_{11}$ received during slot~1 and
uses it to help decode the source message.
The signal of length $t_1 T$ received by the destination during the second slot
can be written as
\begin{equation}\label{eq:y21}
Y_{21}=\sqrt{\phi(\rho)}H_{12}X_{11}+
\sqrt{\alpha_0\rho}H_{02}X_{01}+V_{21}\:.
\end{equation}
Note that~\eqref{eq:y21} can be seen as a Multiple Access Channel (MAC).
In order to recover $X_{11}$ (and consequently $\tilde{Y}_{10}$) 
from~\eqref{eq:y21}, the destination interprets the source contribution as
noise. It succeeds in recovering $\tilde{Y}_{10}$ in case the event
\begin{equation}\label{eq:event_F}
\mathcal{F}=\left\{\omega:t_1\log\left(1+\frac{\phi(\rho)G_{12}(\omega)}{
\alpha_0\rho G_{02}(\omega)+1}\right)>Q(\rho)t_0\right\}
\end{equation}
is realized. We distinguish between three possible cases.\\
{\bf Events $\mathcal{S}$ and $\boldsymbol{\mathcal{F}}$ are realized:} In this
case, the contribution of $X_{11}$ in~\eqref{eq:y21} can be canceled, and the
resulting signal can be written as 
$Y_{21}^{'}=\sqrt{\alpha_0\rho} H_{02}X_{01}+V_{21}$. 
Moreover, it is a straightforward result of~\eqref{eq:y_y} that the conditional
distribution $p_{\tilde{Y}|Y}$ is Gaussian with mean
$\mathbb{E}\left[\tilde{Y}|Y\right]=
\frac{1+\alpha_0\rho G_{01}-\Delta^2(\rho)}{1+\alpha_0\rho G_{01}}Y$ and
variance $\textrm{var}\left(\tilde{Y}|Y\right)=
\frac{\Delta^2(\rho)\left(1+\alpha_0\rho G_{01}-\Delta^2(\rho)\right)}
{1+\alpha_0\rho G_{01}}$. We thus write
\begin{equation}\label{eq:cond_y}
\tilde{Y}_{10}=\frac{1+\alpha_0\rho G_{01}-\Delta^2(\rho)}{1+\alpha_0\rho
G_{01}}Y_{10}+\sqrt{\frac{\Delta^2(\rho)\left(1+\alpha_0\rho
G_{01}-\Delta^2(\rho)\right)}{1+\alpha_0\rho G_{01}}}\tilde{Z}\:,
\end{equation}
where vector $\tilde{Z}$ is AWGN independent of $Y_{10}$ such that each of its
components $\tilde{Z}(i)$ satisfies $\tilde{Z}(i)\sim\mathcal{CN}(0,1)$.
Plugging $Y_{10}=\sqrt{\alpha_0\rho}H_{01}X_{00}+V_{10}$ into~\eqref{eq:cond_y},
it follows that
$$
\tilde{Y}_{10}=\sqrt{\gamma(G_{01},\rho)\alpha_0\rho}H_{01}X_{00}+\tilde{V}_{10}
\:,
$$
where 
$\gamma(G_{01},\rho)=\frac{\left(1+\alpha_0 \rho G_{01}-\Delta^2(\rho)\right)^2}
{\left(1+\alpha_0 \rho G_{01}\right)^2}$ and where vector $\tilde{V}_{10}$ is
AWGN whose components satisfy 
$\tilde{V}_{10}(i)\sim\mathcal{CN}\big(
0,\gamma(G_{01},\rho)+\Delta^2(\rho)\sqrt{\gamma(G_{01},\rho)}
\big)$.
In order to decode the source message, the overall received signal can be
reconstructed as 
$Y_2=\left[Y_{20}^\mathrm{T},\tilde{Y}_{10}^\mathrm{T},
(Y_{21}^{'})^\mathrm{T}\right]^\mathrm{T}$ given by
\begin{equation}\label{eq:y2}
Y_2=\mathbf{H}_{\mathcal{F}}[X_{00}^\mathrm{T},X_{01}^\mathrm{T}]^\mathrm{T}
+\check{V}_{10},
\end{equation}
where
$$
\mathbf{H}_{\mathcal{F}}=\left[
\begin{array}{cc}
\sqrt{\alpha_0\rho}H_{02}\mathbf{I}_{t_0T}& 0\\
\sqrt{\gamma(G_{01},\rho)\alpha_0\rho}H_{01}\mathbf{I}_{t_0T}&
0\\
0& \sqrt{\alpha_0\rho}H_{02}\mathbf{I}_{t_1T}
\end{array}
\right]\:,$$ 
and where $\check{V}_{10}=\left[V_{20}^T,\tilde{V}_{10}^T,V_{21}^T\right]^T$ is
an additive Gaussian noise of zero mean and and whose covariance matrix is given
by
$$
\mathbb{E}[\check{V}_{10}\check{V}_{10}^*]=\left[
\begin{array}{ccc}
\mathbf{I}_{t_0T}& 0& 0\\
0&
\sqrt{\gamma(G_{01},\rho)+\Delta^2(\rho)\sqrt{\gamma(G_{01},\rho)}}\mathbf{I}_{
t_0T } & 0\\
0& 0& \mathbf{I}_{t_1T}
\end{array}
\right]\:.
$$

{\bf Events $\mathcal{S}$ and $\boldsymbol{\overline{\mathcal{F}}}$ are
realized:}
The destination will only be able to use $Y_{20}$, the signal
received during slot~0, to recover the source message. Notice that in such a case, we
get $Y_{20}=\sqrt{\alpha_0\rho}H_{02}X_{00}+V_{20}$.

{\bf Event $\boldsymbol{\overline{\mathcal{S}}}$ is realized:} In this case, 
condition~\eqref{eq:Q_cond_b} is not satisfied and the relay does not
quantize the source message.
This is like the case of a non cooperative transmission.

Finally, the outage probability of the DoQF protocol writes
\begin{align}
P_o(\rho)=P_{o,1}(\rho)+P_{o,2}(\rho)+P_{o,3}(\rho)+P_{o,4}(\rho)\:.
\label{eq:DoQF_o}
\end{align}
where 
\begin{itemize}
\item $P_{o,1}(\rho)$ is the probability that the destination is in outage
\emph{and} that the event $\mathcal{E}$ is realized. We thus get
\begin{equation}
P_{o,1}(\rho)=\mathrm{Pr}[t_0\log(1+\alpha_0\rho G_{02})+t_1\log(1+\alpha_0\rho
G_{02}+\alpha_1\rho G_{12})\leq
R](1-\mathrm{Pr}\left[\overline{\mathcal{E}}\right])\:.
\label{eq:DF_po1}
\end{equation}
where $\mathrm{Pr}\left[\overline{\mathcal{E}}\right]$ is the
probability that the
relay does not succeed in decoding the source message; 
\item $P_{o,2}(\rho)$ is the probability that the destination is in outage and 
that events $\overline{\mathcal{E}}$, $\mathcal{F}$
and $\mathcal{S}$ are realized. We thus have 
\begin{align}
P_{o,2}(\rho)=\mathrm{Pr}\Bigg[&t_1 \log(1+\alpha_0\rho
G_{02})+t_0\log\left(1+\alpha_0\rho
G_{02}+\frac{\gamma(G_{01},\rho)\alpha_0\rho G_{01}}
{\gamma(G_{01},\rho)+\Delta^2(\rho) \sqrt{\gamma(G_{01},\rho)}}\right)
\leq R,\nonumber\\ 
&\overline{\mathcal{E}}, \mathcal{F}, \mathcal{S}\Bigg]\:;\label{eq:DoQF_po3}
\end{align}
\item $P_{o,3}(\rho)$ is the probability that the
destination is in outage and that events $\overline{\mathcal{E}}$, 
$\overline{\mathcal{F}}$ and $\mathcal{S}$ are realized. Therefore we have
\begin{equation}\label{eq:DoQF_po4}
P_{o,3}(\rho)=\mathrm{Pr}[t_0\log(1+\alpha_0\rho
G_{02})\leq R,\overline{\mathcal{E}},
\overline{\mathcal{F}}, \mathcal{S}]\:;
\end{equation}
\item $P_{o,4}(\rho)$ is the probability that
the destination is in outage and that events $\overline{\mathcal{E}}$ and
$\overline{\mathcal{S}}$ are realized. One can easily check that 
\begin{equation}\label{eq:DoQF_po5}
 P_{o,4}(\rho)=\mathrm{Pr}\Big[
 \log(1+\alpha_0\rho G_{02}) \leq R, \overline{\mathcal{E}},
 \overline{\mathcal{S}}
 \Big]\:.
\end{equation}
\end{itemize}

\begin{figure}[h]
\centering\includegraphics[width=12cm]{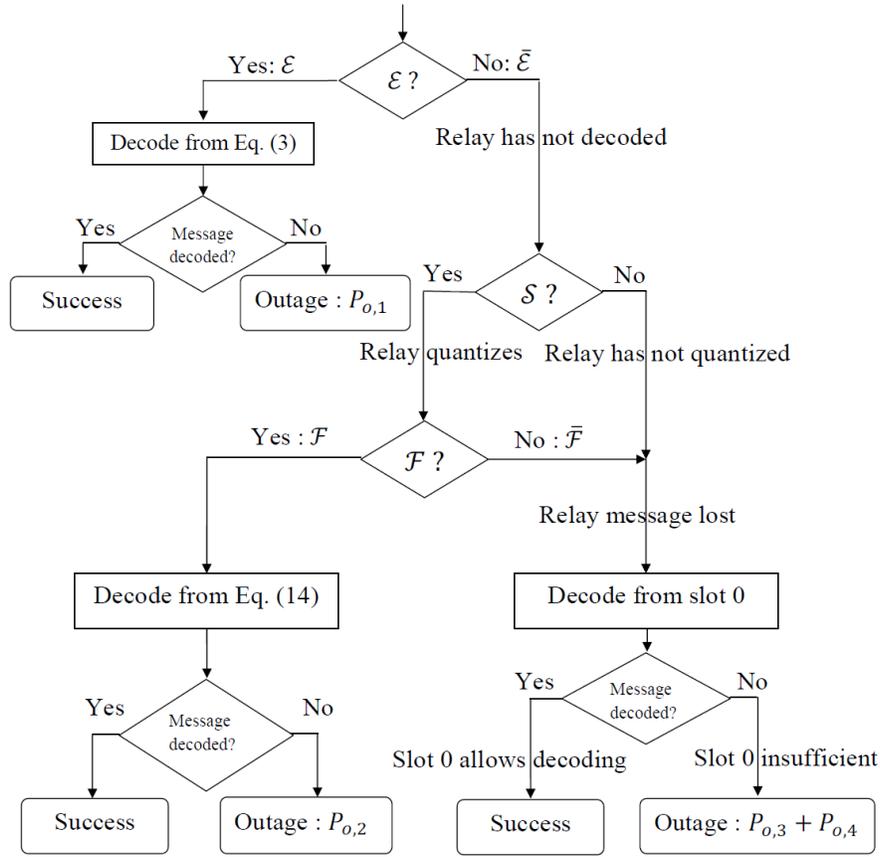}
\caption{Data processing at the destination}
\label{fig:flow_chart}
\end{figure}

In Figure~\ref{fig:flow_chart}, the data processing steps at the destination
node are summarized.
\subsection{On the selection of parameters
$t_0,t_1,\alpha_0,\alpha_1,\phi(\rho),\Delta^2(\rho)$}

In order to complete the definition of the DoQF protocol, we still need to
provide a method for the selection of the relative slots durations
$t_0$, $t_1$, the amplitude factors $\alpha_0$, $\alpha_1$, the
relay power $\phi(\rho)$ and the quantization squared-error $\Delta^2(\rho)$.

We first begin by considering parameters $t_0$, $t_1$, $\alpha_0$, $\alpha_1$,
and $\phi(\rho)$. It is clear that these parameters should be selected such that
the power constraint~\eqref{eq:contrainte} is respected. Recall that the power
constraint~\eqref{eq:contrainte} is a long term constraint which ensures that
the \emph{average} total energy $E_0+E_1$ spent by the network does not exceed
a certain value \emph{i.e.,} $E_0+E_1 \leq \rho T$.
In order to make explicit this power constraint, let us derive the average
energy spent by both the source and the relay to transmit a block of $RT$ nats.
The source transmits the signal
$[\sqrt{\alpha_0\rho}X_{00},\sqrt{\alpha_0\rho}X_{01}]$ spending the energy
$E_0=\alpha_0\rho T$. If the event $\mathcal{E}$ is realized \emph{i.e.,} if the
relay decodes the source message, then the relay transmits the signal
$\sqrt{\alpha_1\rho}X_{11}$ and spends $\alpha_1\rho t_1 T$ Joules. 
If the events $\mathcal{\overline{E}}$ and $\mathcal{S}$ are realized, the relay
transmits $\sqrt{\phi(\rho)}X_{11}$ spending $\phi(\rho) t_1 T$ Joules. As for
the case where the event $\overline{\mathcal{S}}$ is realized, the relay remains
inactive spending no energy. The average energy spent by the relay is thus
$E_1=\alpha_1\rho t_1 T\left(
1-\mathrm{Pr}\left[\overline{\mathcal{E}}\right]
\right)+
\phi(\rho) t_1 T \textrm{Pr}\left[\mathcal{\overline{E}},\mathcal{S}\right]$. 
Putting all pieces together, the power constraint given by~\eqref{eq:contrainte}
can be written as 
\begin{equation}\label{eq:contrainte1}
\alpha_0\rho+\alpha_1\rho
t_1\left(1-\mathrm{Pr}\left[\overline{\mathcal{E}}\right]\right)+\phi(\rho) t_1
\textrm{Pr}\left[\mathcal{\overline{E}},\mathcal{S}\right]\leq \rho\:.
\end{equation}
Parameters $t_0$, $t_1$, $\alpha_0$, $\alpha_1$ and $\phi(\rho)$ should thus be
selected such that constraint~\eqref{eq:contrainte1} is respected. This task
will be addressed in Section~\ref{sec:outage} assuming that the performance
metric is the outage gain, and in Section~\ref{sec:DoQF_DMT} assuming that the
performance metric is the DMT. Note that since the probability
$\textrm{Pr}\left[\mathcal{\overline{E}},\mathcal{S}\right]$ is in
general smaller than $1-\mathrm{Pr}\left[\overline{\mathcal{E}}\right]$ for
sufficiently large values of the SNR
$\rho$, the power $\phi(\rho)$ can be boosted beyond the value $\alpha_1\rho$
without violating the average power constraint given by~\eqref{eq:contrainte1}.

Consider now the quantization squared-error distortion $\Delta^2(\rho)$, and let
us discuss some possible choices for the way $\Delta^2(\rho)$ depends on the
SNR $\rho$. One possible case is to choose $\Delta^2(\rho)$ such that
$\lim_{\rho\to\infty}\Delta^2(\rho)=0$ \emph{i.e.,} \emph{fine quantization} is
achieved at high SNR. This choice will be revealed relevant when the performance
metric is the outage gain (see Section~\ref{sec:outage}). As for the case where
the performance metric is the DMT, we will see in Section~\ref{sec:DoQF_DMT}
that choosing $\lim_{\rho\to\infty}\Delta^2(\rho)=0$ is relevant for some values
of the multiplexing gain $r$, while it is not for other values.

Now that a detailed description of the DoQF has been provided, the rest of the
paper is devoted to the study of the performance of this protocol using two
performance metrics: The outage gain and the DMT.
\section{Outage Probability Analysis of the DoQF Protocol}
\label{sec:outage}

This section is devoted to the outage gain derivation and its minimization
over power and time slot allocation for the DoQF protocol.
\subsection{Notations and Channel Assumptions}
\label{sec:notations_out}

Recall that $H_{ij}$ is the random variable that represents the wireless channel
between nodes $i$ and $j$ of the network ($i,j\in\{0,1,2\}$), 
and that $G_{ij}=|H_{ij}|^2$ designates the power gain of this channel. In this
section, all variables $G_{ij}$ are assumed to have densities $f_{G_{ij}}(x)$
which are right continuous at zero. This assumption is satisfied in particular
by the so-called Rayleigh and Rice channels. Note that except for this mild
assumption, we do not make any assumption on the channels probability
distributions. We denote by $c_{ij}$ the limit $c_{ij}=f_{G_{ij}}(0^+)$ and we
assume that all these limits are positive. For instance, in the Rayleigh case, 
$H_{ij}$ is complex circular Gaussian with zero mean and variance
$\sigma_{ij}^2$. In this case, $G_{ij}$ has the exponential distribution
$f_{G_{ij}}(x)=\sigma_{ij}^{-2}\exp(-x/\sigma_{ij}^2)\mathbf{1}\{x\geq0\}$, and
in particular $c_{ij}=\sigma_{ij}^{-2}$. Here, for any subset $\mathcal{A}$
of $\mathbb{R}$, we denote by $\mathbf{1}\{\mathcal{A}\}$ the indicator function
of the set $\mathcal{A}$. More generally, in the Rician case
$H_{ij}\sim\mathcal{CN}(m_{ij},\sigma_{ij}^2)$ where the mean $m_{ij}$ is not
necessarily zero, the density is given by
$$
f_{G_{ij}}(x)=\frac{1}{\sigma_{ij}^2}\exp\left(-\frac{|m_{ij}|^2+x}{\sigma_{ij}
^2}\right)I_0\left(2\sqrt{x}\frac{|m_{ij}|}{\sigma_{ij}^2}\right)\mathbf{
1}\{x\geq0\}\:,
$$
where $I_0$ is the modified zero order Bessel function of the first kind. As
$I_0(0^+)=1$, we have in this case
$$
c_{ij}=\frac{1}{\sigma_{ij}^2}\exp\left(-\frac{|m_{ij}|^2}{\sigma_{ij}^2}
\right)\:.
$$
In this section, the constants $c_{01}, c_{02}$ and $c_{12}$ are assumed to be
available to the resource allocation unit.

\subsection{Lower Bound on the Outage Gain of Static Half-Duplex Protocols}

Before deriving the outage gain of the proposed DoQF protocol, we first derive a
bound on the outage performance of the wide class of half-duplex static relaying
protocols. This class is indexed in the following by parameters $t_0$,
$\alpha_0$, $\alpha_1$. For each value of these parameters, the class is denoted
by ${\cal P}_{\textrm{HD}}(t_0,\alpha_0,\alpha_1)$ and is defined as the
set of all half-duplex static relaying protocols which satisfy the following.
\begin{itemize}
 \item[-] The source has at its disposal a dictionary of 
 $\left\lfloor e^{RT}\right\rfloor$ codewords. Each codeword
 $X_0=\left[X_{00}^\mathrm{T},X_{01}^\mathrm{T}\right]^\mathrm{T}$ is a vector
 of length $T$ channel uses. 
 \item[-] The source average transmit power
 $\frac{1}{T}\sum_{i=1}^{T}E\left[|X_0(i)|^2\right]$ satisfies the following
 high SNR constraint
 \begin{equation}\label{eq:high_snr_constraint0}
 \lim_{\rho\to\infty}\frac{\frac{1}{T}\sum_{i=1}^{T}
 \mathbb{E}\left[|X_0(i)|^2\right]}{\rho} \leq\alpha_0\:.
 \end{equation}
 \item[-] The relay listens to the source signal during the first $t_0 T$
 channel uses out of the $T$ channel uses which is the duration of the whole
 transmission. The relay has at its disposal a dictionary of codewords $X_{11}$
 of length $(1-t_0)T$ channel uses each. 
 \item[-] During the last $(1-t_0)T$ channel uses of the transmission, the relay
 average transmit power satisfies
 \begin{equation}\label{eq:high_snr_constraint1}
 \lim_{\rho\to\infty}\frac{\frac{1}{(1-t_0)T}\sum_{i=1}^{(1-t_0)T}
 \mathbb{E}\left[|X_{11}(i)|^2\right]}{\rho }\leq\alpha_1\:.
 \end{equation}
\end{itemize}
Note that the above definition does not assume any particular distribution of
the codewords that compose the codebooks of the source and the relay.
Moreover, the definition of the class 
${\cal P}_{\textrm{HD}}(t_0,\alpha_0,\alpha_1)$ imposes no constraints on the
powers transmitted by the nodes for finite values of the SNR $\rho$. Instead,
constraints~\eqref{eq:high_snr_constraint0} and~\eqref{eq:high_snr_constraint1}
restrict only the way the average transmit powers of the source and the
relay behave in the high SNR regime.

\begin{theo}\label{the:genie}
For any static half-duplex relaying protocol from the class 
${\cal P}_{\textrm{HD}}(t_0,\alpha_0,\alpha_1)$, the outage gain
$\xi=\lim_{\rho\to\infty}\rho^2 P_o(\rho)$ is lower-bounded by 
$\xi_{\textrm{CS-HD}}$, where
\begin{align}
\xi_{\mathrm{CS-HD}}=&\frac{c_{02}c_{01}}{\alpha_0^2}\left(\frac{1}{2}+
\frac{\exp(2R)}{4t_0-2}-\frac{t_0\exp(R/t_0)}{2t_0-1}\right)+\nonumber\\
&\frac{c_{02}c_{12}}{\alpha_{0}\alpha_{1}}\left(\frac{1}{2}+
\frac{\exp(2R)}{4t_1-2}-\frac{t_1\exp(R/t_1)}{2t_1-1}\right)\:.\label{eq:genie}
\end{align}
\end{theo}
The proof of Theorem \ref{the:genie} is provided in
Appendix~\ref{app:genie_proof}. The above lower-bound has been derived using the
Cut-Set (CS) bound for Half-Duplex (HD) relay channels, as we will see in
Appendix~\ref{app:genie_proof}. This explains the use of notation
$\xi_{\textrm{CS-HD}}$ with the subscript ($\textrm{CS-HD}$) to designate this
bound.

We now derive and compare the outage gain of the proposed DoQF protocol 
with the above lower-bound.

\subsection{Outage Gain of the DoQF Protocol}

The following theorem characterizes the outage performance of the proposed
protocol at high SNR.
\begin{theo}\label{the:DoQF_out}
Assume that the quantization squared-error distortion $\Delta^2(\rho)$ and
the relay power $\phi(\rho)$ are chosen such that
 \begin{align}
    & \lim_\rho \phi(\rho) = +\infty,\label{eq:lim_rho1_out}\\
    & \lim_{\rho}\frac{\phi(\rho)}{\rho^2} = 0,
    \label{eq:lim_rho2_out}\\
    & \lim_\rho \Delta^2(\rho) = 0,  \label{eq:lim_rho3_out}\\
    & \lim_\rho \left( \phi(\rho)^{t_1}\: \Delta^2(\rho)^{t_0} \right) 
      = +\infty. \label{eq:lim_rho4_out}
 \end{align}
In particular, constraint~\eqref{eq:lim_rho2_out} ensures that the DoQF belongs
to the class ${\cal P}_{\textrm{HD}}(t_0,\alpha_0,\alpha_1)$.
The outage gain $\xi_{\textrm{DoQF}}$ associated with the proposed DoQF protocol
coincides with the lower-bound given by~\eqref{eq:genie}:
$$
\xi_{\mathrm{DoQF}}=\xi_{\mathrm{CS-HD}}\:.
$$
\end{theo}
The proof of Theorem~\ref{the:DoQF_out} is given in
Subsection~\ref{sec:DoQF_out}. 
Theorem~\ref{the:DoQF_out} states that the DoQF protocol is outage-gain-optimal 
in the wide class of half-duplex static relaying protocols.

As a matter of fact, the outage gain associated with the DoQF
 protocol depends on both the quantization error $\Delta^2(\rho)$ and the power
 $\phi(\rho)$ allocated to the relay during slot~1. Theorem~\ref{the:DoQF_out}
 states that it is sufficient to choose $\Delta^2(\rho)$ and $\phi(\rho)$ such
 that constraints~\eqref{eq:lim_rho1_out}-\eqref{eq:lim_rho4_out} are satisfied
 in order for the outage gain of the DoQF to be equal to the lower-bound
 $\xi_{\textrm{CS-HD}}$. 
 The choice $\phi(\rho)=\alpha_1\rho$ is for instance a possible candidate
 for $\phi(\rho)$, provided that $\Delta^2(\rho)$ is chosen such
 that~\eqref{eq:lim_rho3_out} and~\eqref{eq:lim_rho4_out} are satisfied
 \emph{i.e.,} such that
 $\rho^{-\frac{t_1}{t_0}}\stackrel{.}{<}\Delta^2(\rho)\stackrel{.}{<}1$. 
 It is  therefore optimal from an outage gain perspective to let the
 relay transmit at a constant power regardless of whether the source message has
 been decoded with success or not.

\subsection{Proof of Theorem~\ref{the:DoQF_out}}
\label{sec:DoQF_out}

Recall the definition of $P_o(\rho)$ given by~\eqref{eq:DoQF_o} as the outage
probability associated with the DoQF protocol.
In order to prove Theorem~\ref{the:DoQF_out}, we need to show that $\rho^2
P_o(\rho)$ converges as $\rho\to\infty$ and to derive the outage gain
$\xi_{\textrm{DoQF}}$ given by 
$\xi_{\textrm{DoQF}}=\lim_{\rho\to\infty}\rho^2 P_o(\rho)$.
According to~\eqref{eq:DoQF_o},
$P_o(\rho)=P_{o,1}(\rho)+P_{o,2}(\rho)+P_{o,3}(\rho)+P_{o,4}(\rho)$,
where $P_{o,1}(\rho)$, $P_{o,2}(\rho)$, $P_{o,3}(\rho)$ and $P_{o,4}(\rho)$ are
defined by~\eqref{eq:DF_po1}, \eqref{eq:DoQF_po3}, \eqref{eq:DoQF_po4} and
\eqref{eq:DoQF_po5} respectively. Therefore, we need to first compute the
limits
$\lim_{\rho\to\infty}\rho^2 P_{o,1}(\rho)$, 
$\lim_{\rho\to\infty}\rho^2 P_{o,2}(\rho)$,
$\lim_{\rho\to\infty}\rho^2 P_{o,3}(\rho)$ and 
$\lim_{\rho\to\infty}\rho^2 P_{o,4}(\rho)$
in order to obtain the outage gain $\xi_{\textrm{DoQF}}$.
It has been proved in~\cite{outage_gain} that
\begin{equation}\label{eq:DF_po1_lim}
 \lim_{\rho\to\infty}\rho^2 P_{o,1}(\rho)=
\frac{c_{02}c_{12}}{\alpha_0\alpha_1}\int_{\mathbb{R}_+^2}
\mathbf{1}\{t_0\log(1+u)+t_1\log(1+u+v)\leq R\}du dv\:,
\end{equation}
where $c_{01}$ and $c_{12}$ has been defined in
Subsection~\ref{sec:notations_out} as $c_{01}=f_{G_{01}}(0+)$ and
$c_{12}=f_{G_{12}}(0+)$ respectively. The steps of the proof
that~\eqref{eq:DF_po1_lim} holds are very similar to the steps given below for
the derivation of $\lim_{\rho\to\infty}\rho^2 P_{o,2}(\rho)$. Refer to the
definition of $P_{o,2}(\rho)$ given by~\eqref{eq:DoQF_po3} as
\begin{align}
P_{o,2}(\rho)=\mathrm{Pr}\Bigg[&
t_1 \log(1+\alpha_0\rho G_{02})+t_0\log\left(1+\alpha_0\rho G_{02}+
\frac{\gamma(G_{01},\rho)\alpha_0\rho G_{01}}
{\gamma(G_{01},\rho)+\Delta^2(\rho)\sqrt{\gamma(G_{01},\rho)}}\right)<R,
\nonumber\\
&\overline{\mathcal{E}},\mathcal{F},\mathcal{S}\Bigg]\:,\label{eq:po2_temp}
\end{align}
where $\gamma(G_{01},\rho)=\frac{\left(1+\alpha_0\rho
G_{01}-\Delta^2(\rho)\right)^2}{\left(1+\alpha_0\rho G_{01}\right)^2}$. Plugging
the definitions of events $\mathcal{E}$, $\mathcal{S}$ and $\mathcal{F}$ given
respectively by~\eqref{eq:event_E}, \eqref{eq:event_S} and~\eqref{eq:event_F}
into~\eqref{eq:po2_temp} leads to
\begin{align*}
 P_{o,2}(\rho)=&\int_{\mathbb{R}_+^3}
         \mathbf{1}\left\{t_1\log(1+\alpha_0\rho x)+
         t_0\log\left(1+\alpha_0\rho x+
         \frac{\gamma(y,\rho)\alpha_0\rho y}
	 {\gamma(y,\rho)+\Delta^2(\rho)\sqrt{\gamma(y,\rho)}}\right)
         \leq R\right\}\\
         &\times\mathbf{1}\left\{t_0\log(1+\alpha_0\rho y)\leq R\right\}
         \mathbf{1}\left\{1+\alpha_0 \rho y>\Delta^2(\rho)\right\}\\
         &\times\mathbf{1}\left\{t_1\log\left(1+\frac{\phi(\rho)z}
         {1+ \alpha_0 \rho x}\right)> t_0 Q(\rho)\right\}
         f_{G_{02}}(x)f_{G_{01}}(y)f_{G_{12}}(z)dx dy dz\:,
\end{align*}
By making the change of variables $u=\alpha_0\rho x$
and $v=\alpha_0\rho y$ we obtain
\begin{align}
  \rho^2 P_{o,2}(\rho)=\frac{1}{\alpha_0^2}&\int_{\mathbb{R}_+^3}
         \mathbf{1}\left\{t_1\log(1+u)+t_0\log\left(1+u+      
         \frac{\gamma(v,\rho)v}
         {\gamma(v,\rho)+\Delta^2(\rho)\sqrt{\gamma(v,\rho)}}
         \right)\leq R\right\}\nonumber\\
         &\times\mathbf{1}\left\{t_0\log(1+v)\leq R\right\}
          \mathbf{1}\left\{1+v>\Delta^2(\rho)\right\}\nonumber\\
         &\times\mathbf{1} \left\{t_1\log\left(1+\frac{\phi(\rho)z}
         {1+ u}\right)>t_0 Q(\rho)\right\}
f_{G_{02}}\left(\frac{u}{\alpha_0\rho}\right)f_{G_{01}}\left(\frac{v}{
\alpha_0\rho}\right)
         f_{G_{12}}(z)du dv dz\:.\label{eq:DoQF_rho_po3}
\end{align}
Since $Q(\rho)=\log\left(K/\Delta^2(\rho)\right)$ as given by~\eqref{eq:Q}, 
it is possible and useful to write the last indicator as follows.
\begin{equation}\label{eq:z_theta}
\mathbf{1}\left\{t_1\log\left(1+\frac{\phi(\rho)z}
{1+ u}\right)> t_0 Q(\rho)\right\}=
\mathbf{1}\left\{z>(1+u)\theta(\rho)\right\}\:,
\end{equation}
where 
\begin{equation}\label{eq:theta}
\theta(\rho)=\frac{K^{\frac{t_0}{t_1}}}{\phi(\rho)
\left(\Delta^2(\rho)\right)^{\frac{t_0}{t_1}}}-\frac{1}{\phi(\rho)}\:.
\end{equation}
Define the function $\Phi(u,v,z,\rho)$ as the integrand in the rhs
of~\eqref{eq:DoQF_rho_po3} and let $\mathcal{C}$ be the compact subset of
$\mathbb{R}_+^2$ defined as
$\mathcal{C}=\Bigg\{(u,v)\in\mathbb{R}_+^2,
t_1\log(1+u)+t_0\log\left(1+u+\frac{\gamma(v,\rho)v}        
{\gamma(v,\rho)+\Delta^2(\rho)\sqrt{\gamma(v,\rho)}}
\right)\leq R, t_0\log(1+v)\leq$ $R\Bigg\}$.
As $f_{G_{02}}$ and $f_{G_{01}}$ are right continuous at zero, then the
function $(u,v)\mapsto f_{G_{02}}\left(\frac{u}{\alpha_0\rho}\right)f_{G_{01}}
\left(\frac{v}{\alpha_0\rho}\right)$ is bounded on $\mathcal{C}$
for $\rho$ large enough \emph{i.e.}, there exist $\rho_0>0$ and $M>0$ such
that $\forall \rho\geq\rho_0$, $f_{G_{02}}\left(\frac{u}{\alpha_0\rho}\right)
f_{G_{01}}\left(\frac{v}{\alpha_0\rho}\right)\leq M$. It is straightforward to
verify that the following inequalities hold for all $\rho\geq\rho_0$:
\begin{align*}
 \Phi(u,v,z,\rho)\leq & M \times \mathbf{1}\left\{t_1\log(1+u)+t_0\log
 \left(1+u+\frac{\gamma(v,\rho)v}
 {\gamma(v,\rho)+\Delta^2(\rho)\sqrt{\gamma(v,\rho)}}\right)\leq
         R\right\}\\
         &\times\mathbf{1}\left\{t_0\log(1+v)\leq R\right\}
         \mathbf{1}\left\{1+v>\Delta^2(\rho)\right\}\\
         &\times\mathbf{1}\left\{z>(1+u)\theta(\rho)\right\}
         f_{G_{12}}(z)\\
         \leq & M \times\mathbf{1}\left\{\log(1+u)\leq R\right\}
         \times\mathbf{1}\left\{t_0\log(1+v)\leq R\right\}f_{G_{12}}(z)\:.
\end{align*}
The rhs of the latter inequality is an integrable function on $\mathbb{R}_+^3$
and it does not depend on $\rho$. Therefore, we can apply Lebesgue's Dominated
Convergence Theorem (DCT) in order to compute $\lim_{\rho\to\infty}\rho^2
P_{o,2}(\rho)$ in~\eqref{eq:DoQF_rho_po3}. Note first that
$\lim_{\rho\to\infty}\Delta^2(\rho)=0$,
$\lim_{\rho\to\infty}\frac{\gamma(v,\rho)}{\gamma(v,\rho)+\Delta^2(\rho)
\sqrt{\gamma(v,\rho)}} =1$ and $\lim_{\rho\to\infty}\theta(\rho)=0$ due to
assumptions~\eqref{eq:lim_rho1_out}-~\eqref{eq:lim_rho4_out}. 
After some algebra, we can easily show that the following result holds.
\begin{equation}\label{eq:DoQF_po3_lim}
 \lim_{\rho\to\infty}\rho^2 P_{o,2}(\rho)=
 \frac{c_{02}c_{01}}{\alpha_0^2}\int_{\mathbb{R }_+^2}
 \mathbf{1}\left\{t_1\log(1+u)+t_0\log\left(1+u+v\right)\leq R\right\}
 du dv\:.
\end{equation}

We now prove that $\lim_{\rho\to\infty}\rho^2 P_{o,3}(\rho)=0$. First, recall 
that
$P_{o,3}(\rho)=\mathrm{Pr}[t_0\log(1+\alpha_0\rho
G_{02})<R,\overline{\mathcal{E}},
\overline{\mathcal{F}}, \mathcal{S}]$. Plugging the definition of events
$\mathcal{E}$, $\mathcal{S}$ and $\mathcal{F}$ from~\eqref{eq:event_E},
\eqref{eq:event_S} and~\eqref{eq:event_F} respectively into the latter equation
leads to
\begin{align*}
 P_{o,3}(\rho)=&\int_{\mathbb{R}_+^3}\mathbf{1}\left\{t_0\log(1+\alpha_0\rho
x)\leq
         R\right\}
         \mathbf{1}\left\{t_0\log(1+\alpha_0\rho y)\leq R\right\}
         \mathbf{1}\left\{1+\alpha_0 \rho y>\Delta^2(\rho)\right\}\\
         &\times\mathbf{1}\left\{t_1\log\left(1+\frac{\phi(\rho)z}
         {1+ \alpha_0 \rho x}\right)\leq t_0 Q(\rho)\right\}
         f_{G_{02}}(x)f_{G_{01}}(y)f_{G_{12}}(z)dx dy dz\:,
\end{align*}

Defineing $u=\alpha_0 \rho x$ and $v=\alpha_0 \rho y$, we get
\begin{align*}
 P_{o,3}(\rho)=\frac{1}{\alpha_0^2\rho^2}&\int_{\mathbb{R}_+^3}
	 \mathbf{1}\left\{t_0\log(1+u)\leq R\right\}
         \mathbf{1}\left\{t_0\log(1+v)\leq R\right\}
         \mathbf{1}\left\{1+v>\Delta^2(\rho)\right\}\\
         &\times\mathbf{1}\left\{t_1\log\left(1+\frac{\phi(\rho)z}
         {1+ u}\right)\leq t_0 Q(\rho)\right\}
         f_{G_{02}}\left(\frac{u}{\alpha_0\rho}\right)
         f_{G_{01}}\left(\frac{v}{\alpha_0\rho}\right)f_{G_{12}}(z)du dv dz\:.
\end{align*}
As we did in~\eqref{eq:z_theta}, we write the last indicator as follows.
$$
\mathbf{1}\left\{t_1\log\left(1+\frac{\phi(\rho)z}
{1+ u}\right)\leq t_0
Q(\rho)\right\}=\mathbf{1}\left\{z\leq(1+u)\theta(\rho)\right\}\:,
$$
where $\theta(\rho)$ is defined by~\eqref{eq:theta}. 
In analogy with the approach we used to compute $\lim_{\rho\to\infty}\rho^2
P_{o,2}(\rho)$, we define $\mathcal{C}_1$ as the compact subset of
$\mathbb{R}_+^3$ satisfying
$\mathcal{C}_1=\big\{(u,v,z)\in\mathbb{R}_+^3$,
$t_0\log(1+u)\leq R$, $t_0\log(1+v)\leq R$, $z\leq (1+u)\theta(\rho)\big\}$. 
Next, we use the fact that $f_{G_{02}}$, $f_{G_{01}}$ and $f_{G_{12}}$ are right
continuous at zero, along with $\lim_{\rho\to\infty}\theta(\rho)=0$, to show
that the function
$(u,v,z)\mapsto f_{G_{02}}\left(\frac{u}{\alpha_0\rho}\right)f_{G_{01}}
\left(\frac{v}{\alpha_0\rho}\right)f_{G_{12}}(z)$ is bounded on $\mathcal{C}_1$
for $\rho$ large enough \emph{i.e.}, there exist $\rho_1>0$ and $M_1>0$ such
that $\forall \rho\geq\rho_1$, $f_{G_{02}}\left(\frac{u}{\alpha_0\rho}\right)
f_{G_{01}}\left(\frac{v}{\alpha_0\rho}\right)f_{G_{12}}(z)\leq M_1$.
It follows that the following inequalities hold for all $\rho\geq\rho_1$:
\begin{align*}
 \rho^2 P_{o,3}(\rho)&\leq
\frac{M_1}{\alpha_0^2}\int_{\mathbb{R}_{+}^2}\mathbf{1}
        \left\{1+u \leq e^{\frac{R}{t_0}}\right\}
         \mathbf{1}\left\{z\leq (1+u)\theta(\rho)\right\}
         du dz\\
        &\leq \frac{M_1}{\alpha_0^2}\int_{\mathbb{R}_{+}}\mathbf{1}\left\{z\leq
         e^{\frac{R}{t_0}}\theta(\rho)\right\}dz 
	\leq \frac{M_1}{\alpha_0^2}\int_{0}^{e^{\frac{R}{t_0}}\theta(\rho)}dz
        =\frac{M_1}{\alpha_0^2} e^{\frac{R}{t_0}} \theta(\rho)\:.
\end{align*}
Now since $\lim_{\rho\to\infty}\theta(\rho)=0$ due to
assumptions~\eqref{eq:lim_rho1_out}-\eqref{eq:lim_rho4_out}, it follows that 
$\lim_{\rho\to\infty}\rho^2 P_{o,3}(\rho)=0$. We can prove in the same way and
without difficulty that
$\lim_{\rho\to\infty}\rho^2 P_{o,4}(\rho)=0$.

Putting all pieces together, 
\begin{align}\label{eq:DoQF_po_lim}
 \lim_{\rho\to\infty}\rho^2 P_o = 
 &\lim_{\rho\to\infty}\rho^2 P_{o,1}(\rho)
 +\lim_{\rho\to\infty}\rho^2 P_{o,2}(\rho)
 +\lim_{\rho\to\infty}\rho^2 P_{o,3}(\rho)
 +\lim_{\rho\to\infty}\rho^2 P_{o,4}(\rho)\nonumber\\
 =&\frac{c_{02}c_{12}}{\alpha_0\alpha_1}\int_{\mathbb{R}_+^2}
\mathbf{1}\{t_0\log(1+u)+t_1\log(1+u+v)\leq R\}du dv\nonumber\\
+&\frac{c_{02}c_{01}}{\alpha_0^2}\int_{\mathbb{R }_+^2}
 \mathbf{1}\left\{t_1\log(1+u)+t_0\log\left(1+u+v\right)\leq R\right\}
 du dv\:.
\end{align}
The remaining task is to prove that the rhs of~\eqref{eq:DoQF_po_lim} is equal
to the rhs of~\eqref{eq:genie}.
This can be done by making the change of variables $x=\log(1+u)$ and
$y=\log\left(1+\frac{v}{1+u}\right)$ in~\eqref{eq:DoQF_po_lim}. The
details of the proof can be found in~\cite{outage_gain}. The proof of
Theorem~\ref{the:DoQF_out} is thus complete.

\subsection{Power and Time Optimization}
\label{sec:opt}
Our aim in this subsection is to derive the slots relative durations $t_0$,
$t_1$ and the power allocation parameters $\alpha_0$, $\alpha_1$ that minimizes
the outage gain $\xi_{\textrm{DoQF}}$ associated with the DoQF protocol. This
minimization should be done subject to the power constraint given
by~\eqref{eq:contrainte1}. Let us examine the above constraint when the SNR
$\rho$ tends to infinity. We first divide the two sides of the power constraint
by $\rho$, which leads to 
$\alpha_0+\alpha_1
t_1\left(1-\mathrm{Pr}\left[\overline{\mathcal{E}}\right]\right)+\frac{
\phi(\rho) } { \rho } t_1
\textrm{Pr}\left[\mathcal{\overline{E}},\mathcal{S}\right]
\leq 1$, where $\mathrm{Pr}[\overline{\mathcal{E}}]=
\mathrm{Pr}\left[t_0\log(1+\alpha_0\rho G_{01})\leq R\right]$.
It is useful to write the term $\frac{\phi(\rho)}{\rho} t_1
\textrm{Pr}\left[\mathcal{\overline{E}},\mathcal{S}\right]$ in the lhs
of the above inequality as $t_1 \frac{\phi(\rho)}{\rho^2} \rho
\textrm{Pr}\left[\mathcal{\overline{E}},\mathcal{S}\right]$.
Recall that due to~\eqref{eq:lim_rho2_out}, $\phi(\rho)$ is chosen such that
$\lim_{\rho\to\infty}\frac{\phi(\rho)}{\rho^2}=0$. Furthermore, it is
straightforward to check that $\rho
\textrm{Pr}\left[\mathcal{\overline{E}},\mathcal{S}\right]$ is upper-bounded for
any $\rho\in\mathbb{R}_+$. Indeed, $\lim_{\rho\to\infty}\rho
\textrm{Pr}\left[\mathcal{\overline{E}},\mathcal{S}\right]$ is a constant.
Putting all pieces together, the power constraint at high SNR writes as
$\alpha_0+t_1\alpha_1\leq1\:.$
Note that this constraint is not convex in $\alpha_0,\alpha_1,t_1$ because
the function $(\alpha_1,t_1)\mapsto \alpha_1 t_1$ is not. It will be
convenient to replace it with a convex constraint by making the change of
variables $\beta_0=\alpha_0$ and $\beta_1=\alpha_1 t_1$. 
With these new variables, the power constraint becomes
\begin{equation}\label{eq:contrainte3}
\beta_0+\beta_1\leq1\:,
\end{equation}
and the outage gain $\xi_{\textrm{DoQF}}$ given by~\eqref{eq:DoQF_o} writes
\begin{align}
\xi_{\textrm{DoQF}}(t_1,\beta_0,\beta_1)
=&\frac{c_{02}c_{01}}{\beta_0^2}\left(\frac{1}{2}+
\frac{\exp(2R)}{2-4t_1}-\frac{(1-t_1)}{1-2t_1}\exp\left(\frac{R}{1-t_1}
\right)\right)+\nonumber\\
&\frac{c_{02}c_{12}t_1}
{\beta_{0}\beta_{1}}\left(\frac{1}{2}+\frac{\exp(2R)}{4t_1-2}-
\frac{t_1}{2t_1-1}\exp\left(\frac{R}{t_1}\right)\right)
\label{eq:out_gain_convex} \:.
\end{align}
Using the same arguments of~\cite{outage_gain}, it can be shown that the
function $\xi_{\textrm{DoQF}}(t_1,\beta_0,\beta_1)$ is convex on the domain
$(0,1)\times(0,\infty)^2$.
The outage probability minimization at high SNR reduces thus to minimizing
$\xi_{\textrm{DoQF}}(t_1,\beta_0,\beta_1)$ given the
constraint~\eqref{eq:contrainte3}. This in turn reduces to minimizing
$\xi_{\textrm{DoQF}}$ on the line segment of $\mathbb{R}_+^2$ defined
by $\beta_0+\beta_1=1$ \emph{i.e.,} the constraint~\eqref{eq:contrainte3} is met
with equality. The function $\xi_{\textrm{DoQF}}(t_1,\beta_0,1-\beta_0)$ defined
on the open square $(0,1)^2$ is convex as it coincides with the restriction of
$\xi_{\textrm{DoQF}}(t_1,\beta_0,\beta_1)$ to a line segment. Furthermore, it
is clear that $\xi_{\textrm{DoQF}}(t_1,\beta_0,1-\beta_0)$ goes to infinity on
the frontier of $(0,1)^2$. Therefore, the minimum is in the interior
of $(0,1)^2$, and can be obtained easily, for instance by a suitable descent
method~\cite{convex_opt}.

\section{DMT Analysis of the DoQF Protocol}\label{sec:DoQF_DMT}

This section is devoted to the derivation of the DMT of the proposed DoQF
protocol.

\subsection{Channel Assumptions}

In this section, the wireless channels between the different nodes of
the network are assumed to be Rayleigh distributed. This assumption
is to be compared with the mild assumptions involved in the derivation
of the outage gain in Section~\ref{sec:outage}, and which apply to a large
class of channel distributions, including Rayleigh and Rice channels.
Finally, the transmission rate is assumed to be a function of the SNR $\rho$ and
to satisfy $R=R(\rho)\stackrel{.}{=}r \log\rho$, in accordance
with~\eqref{eq:DMT_def}. 

Before proceeding with the derivation of the DMT of the proposed DoQF protocol,
we should first select the way the quantization squared-error $\Delta^2(\rho)$
and the relay power $\phi(\rho)$ depend on the SNR $\rho$.

\subsection{On the Selection of $\Delta^2(\rho)$ and $\phi(\rho)$ from a DMT
Perspective}

The outage probability $P_o$ associated with the DoQF protocol and defined
by~\eqref{eq:DoQF_o} depends on the quantization squared-error distortion
$\Delta^2(\rho)$ and on the power $\phi(\rho)$ allocated to the relay during
slot~1. Consequently, the DMT associated with the DoQF depends likewise on these
two parameters. In Section~\ref{sec:outage}, parameters $\Delta^2(\rho)$ and
$\phi(\rho)$ were chosen such that
constraints~\eqref{eq:lim_rho1_out}-\eqref{eq:lim_rho4_out} are satisfied. 
Moreover, it was shown that this choice is optimal from an outage gain
perspective. In the current section, we are interested in choices of
$\Delta^2(\rho)$ and $\phi(\rho)$ that are relevant from a DMT perspective.
In the sequel, we compute the DMT of the DoQF assuming that
\begin{equation}\label{eq:delta_dmt}
 \Delta^2(\rho)\stackrel{.}{=}\rho^\delta\:,
\end{equation} 
where parameter $\delta$ will be fixed later.

As for the power $\phi(\rho)$, it should be chosen such that the best
possible DMT is achieved by the protocol without violating the power constraint
given by~\eqref{eq:contrainte1}. Since we are evaluating the performance of
the DoQF protocol from a DMT perspective, this power constraint should be
examined in the asymptotic regime where $\rho$ tends to infinity. We remind that 
the term $\textrm{Pr}\left[\mathcal{\overline{E}},\mathcal{S}\right]$
in~\eqref{eq:contrainte1}, to begin with, denotes the
probability that events $\overline{\mathcal{E}}$ and $\mathcal{S}$ are realized
\emph{i.e.,} 
$\textrm{Pr}\left[\mathcal{\overline{E}},\mathcal{S}\right]=\mathrm{Pr}\left[
t_0\log(1+\alpha_0\rho G_{01})\leq R(\rho),1+\alpha_0\rho
G_{01}>\Delta^2(\rho)\right]$. 
It is straightforward to show that
$\textrm{Pr}\left[\mathcal{\overline{E}},\mathcal{S}\right]\stackrel{.}{=}
\rho^{-\left(1-r/t_0\right)^+}$ provided that 
$\delta\leq 1-\left(1-\frac{r}{t_0}\right)^+$. We will see later on that 
$\delta\leq 1-\left(1-\frac{r}{t_0}\right)^+$ is the relevant choice
for $\delta$ from a DMT point of view, and is thus assumed in the sequel.
Plugging this result into~\eqref{eq:contrainte1}, the power constraint
can be written in the asymptotic regime as
\begin{equation}\label{eq:contrainte_asym}
\phi(\rho)\stackrel{.}{\leq}\rho^{1+\left(1-r/t_0\right)^+}\:.
\end{equation}
In order for the DoQF protocol to achieve the best possible DMT, we should
choose $\phi(\rho)$ such that constraint~\eqref{eq:contrainte_asym} is met with
equality. From now on, $\phi(\rho)$ is thus assumed to satisfy
$\phi(\rho)\stackrel{.}{=}\rho^{1+\left(1-r/t_0\right)^+}$.

Note that if we choose $\delta$ such that $\delta<0$, then $\Delta^2(\rho)$
and $\phi(\rho)$ as given by~\eqref{eq:delta_dmt}
and~\eqref{eq:contrainte_asym} also satisfy
constraints~\eqref{eq:lim_rho1_out}-\eqref{eq:lim_rho4_out}. However, choosing
$\Delta^2(\rho)$ and $\phi(\rho)$ that satisfy at the same time constraints
~\eqref{eq:delta_dmt}-\eqref{eq:contrainte_asym} and
constraints~\eqref{eq:lim_rho1_out}-\eqref{eq:lim_rho4_out} does not
necessarily yield the best DMT performance of the protocol, as we will see.

\subsection{DMT of the DoQF protocol}

Now that the power $\phi(\rho)$ allocated to the relay during slot~1 has been
determined, the outage probability of the DoQF protocol depends only on
parameters $t_0$ and $\delta$. Therefore, the DMT associated with the DoQF
protocol should be defined first for fixed values of $t_0$ and $\delta$.
We denote by $d(t_0,\delta,r)$ this DMT which is given by
\begin{equation}\label{eq:DoQF_DMT_o}
d(t_0,\delta,r)=-\lim_{\rho\to\infty}\frac{\log P_o(\rho)}{\log\rho}\:,
\end{equation}
where $P_o(\rho)$ is the outage probability associated with the protocol.
We define the final DMT of DoQF as
\begin{equation}\label{eq:DoQF_DMT_sup}
d_{\mathrm{{DoQF}}}^*(r)=\sup_{t_0,\delta}d(t_0,\delta,r)\:,
\end{equation} 
where the maximization is done with respect to parameters $t_0$ and $\delta$.
Define $t_{0,{\mathrm{DoQF}}}^*(r)$ and
$\delta_{\mathrm{DoQF}}^*(r)$ as the argument of the supremum
in~\eqref{eq:DoQF_DMT_sup}. Theorem~\ref{the:DoQF_DMT} provides the closed-form
expression of the final DMT of the DoQF.
\begin{theo}\label{the:DoQF_DMT}
Assume that the quantization squared-error distortion chosen by the relay
satisfies $\Delta^2(\rho)\stackrel{.}{=}\rho^\delta$.
The DMT $d_{\mathrm{DoQF}}^*(r)$ associated with the DoQF
protocol is given by
\begin{equation}\label{eq:DoQF_dr}
d_{\mathrm{DoQF}}^*(r)=\left\{\begin{array}{ll}
2(1-r)^+ &\mathrm{ for}\: r\leq\frac{1}{4}\\
2-\frac{r}{1-v^*(r)} &\mathrm{ for}\: \frac{1}{4}<r\leq
\frac{2(\sqrt{5}-1)}{9-\sqrt{5}}\\
2-\frac{2}{3-\sqrt{5}}r &\mathrm{ for}\:
\frac{2(\sqrt{5}-1)}{9-\sqrt{5}}<r\leq \frac{\sqrt{5}-1}{\sqrt{5}+1}\\
(2-r)(1-r) &\mathrm{ for}\: r>\frac{\sqrt{5}-1}{\sqrt{5}+1}
\end{array}\right.\:,
\end{equation}
where $v^*(r)$ is the unique solution in
$\left[\frac{1}{2},\frac{2}{\sqrt{5}+1}\right]$ to the
following equation.
\begin{equation}\label{eq:equation_t0}
2(1+r)v^3-(4+5r)v^2+2(1+4r)v-4r=0\:.
\end{equation}
Moreover, the optimal value of $t_0$ as function of $r$ that allows to achieve
the DMT $d_{\mathrm{DoQF}}^*(r)$ is given by
\begin{align}\label{eq:DoQF_t0}
t_{0,{\mathrm{DoQF}}}^*(r)=\left\{\begin{array}{ll}
\frac{1}{2}   &\mathrm{ for}\: 0 \leq r \leq \frac{1}{4}\\
v^*(r)      &\mathrm{ for}\: \frac{1}{4}<r\leq
\frac{2(\sqrt{5}-1)}{9-\sqrt{5}}\\
\frac{2}{\sqrt{5}+1} &\mathrm{ for}\:
\frac{2(\sqrt{5}-1)}{9-\sqrt{5}}<r\leq \frac{\sqrt{5}-1}{\sqrt{5}+1}\\
\frac{1}{2-r} &\mathrm{ for}\: r>\frac{\sqrt{5}-1}{\sqrt{5}+1}
\end{array}\right.\:,
\end{align}
and the optimal value of $\delta$  as function of $r$ that allows to achieve
the DMT $d_{\mathrm{DoQF}}^*(r)$ can be written as
\begin{align}\label{eq:DoQF_delta}
 \delta_{\mathrm{DoQF}}^*(r)=\left\{\begin{array}{ll}
0   &\mathrm{ for}\: 0 \leq r \leq \frac{1}{4}\\
4\frac{r}{v^*(r)}+2(r+1)v^*(r)-2-5r  &\mathrm{ for}\: \frac{1}{4}<r\leq
\frac{2(\sqrt{5}-1)}{9-\sqrt{5}}\\
\frac{r}{t_{0,{\mathrm{DoQF}}}^*(r)} &\mathrm{ for}\:
r>\frac{2(\sqrt{5}-1)}{9-\sqrt{5}}\:.
\end{array}\right.\:.
\end{align}
\end{theo}

The proof of Theorem~\ref{the:DoQF_DMT} is given in Subsection \ref{sec:DoQF_DMT_deriv}.

From this theorem, we can see that the MISO upper-bound
is reached by the DoQF for $r<0.25$, and that the DMT of the protocol deviates
from the MISO bound for $r>0.25$. 

The DMT of (non-orthogonal) DF in the general multiple-relay case has been
derived in~\cite{DMG}. Denote by $P_{o,\mathrm{DF}}(\rho)$ the outage
probability associated with the DF protocol. The DMT of DF for fixed values of
$t_0$ can thus be defined as
\begin{equation}\label{eq:DF_DMT_fix}
d(t_0,r)=-\lim_{\rho\to\infty}\frac{\log
P_{o,\textrm{DF}}(\rho)}{\log\rho}\:,
\end{equation}
and the final DMT of the protocol as $d_{\mathrm{DF}}^*(r)=\sup_{t_0}d(t_0,r)$.
The closed-form expression of $d_{\mathrm{DF}}^*(r)$ in the case of a single
relay is reproduced here by 
\begin{equation}\label{eq:DF_DMT}
d_{\textrm{DF}}^*(r)=\left\{\begin{array}{ll}
2-\frac{2}{3-\sqrt{5}}r &\mathrm{ for}\: 0 \leq r \leq
\frac{\sqrt{5}-1}{\sqrt{5}+1}\\
(2-r)(1-r) &\mathrm{ for}\: \frac{\sqrt{5}-1}{\sqrt{5}+1}<r\leq 1\:.
\end{array}\right.
\end{equation}
Moreover, the optimal value of $t_0$, as function of $r$, that allows to
achieve this DMT is given by
\begin{equation}\label{eq:DF_t0}
t_{0,{\mathrm{DF}}}^*(r)=\left\{\begin{array}{ll}
\frac{2}{\sqrt{5}+1} &\mathrm{ for}\: 0 \leq r \leq
\frac{\sqrt{5}-1}{\sqrt{5}+1}\\
\frac{1}{2-r} &\mathrm{ for}\: \frac{\sqrt{5}-1}{\sqrt{5}+1}<r\leq 1\:.
\end{array}\right.
\end{equation}
We note that the DMT of the DoQF is larger than that of DF on the range of
multiplexing gains $r\leq \frac{2(\sqrt{5}-1)}{9-\sqrt{5}}$.
But for higher values of $r$, quantization at the relay can no more improve the
DMT of the DoQF which becomes equal to the DMT of the DF.

In order to obtain the best possible DMT as given by
Theorem~\ref{the:DoQF_DMT}, we allowed parameters $t_0$ and $\delta$ to depend
on the multiplexing gain~$r$. This additional degree of freedom will not change
the fact that the proposed DoQF protocol is static. Indeed, parameters $t_0$ and
$\delta$ in our model do not depend on any channel coefficients.

\subsection{Proof of Theorem~\ref{the:DoQF_DMT}}
\label{sec:DoQF_DMT_deriv}

The outage probability associated with the DoQF protocol was given
by~\eqref{eq:DoQF_o} as
\begin{equation}\label{eq:po_temp}
P_o(\rho)=P_{o,1}(\rho)+P_{o,2}(\rho)+P_{o,3}(\rho)+P_{o,4}(\rho)\:,
\end{equation}
where probabilities $P_{o,1}(\rho)$, $P_{o,2}(\rho)$, $P_{o,3}(\rho)$ and
$P_{o,4}(\rho)$ are
respectively defined by~\eqref{eq:DF_po1}, \eqref{eq:DoQF_po3},
\eqref{eq:DoQF_po4} and~\eqref{eq:DoQF_po5}. Inserting~\eqref{eq:po_temp} into
the definition of the DMT $d(t_0,\delta,r)$ given by~\eqref{eq:DoQF_DMT_o} leads
to
\begin{align}
d(t_0,\delta,r)=&
-\lim_{\rho\to\infty}\frac{\log\left(
P_{o,1}(\rho)+P_{o,2}(\rho)+P_{o,3}(\rho)+P_{o,4}(\rho)
\right)}{\log\rho}\nonumber\\
=&\min\left\{
d_1(t_0,r),d_2(t_0,\delta,r),d_3(t_0,\delta,r),d_4(t_0,\delta,r)
\right\}\:,\label{eq:dr_min}
\end{align}
where
\begin{equation}\label{eq:di_temp}
d_i(t_0,\delta,r)=-\lim_{\rho\to\infty}\frac{\log P_{o,i}(\rho)}{\log\rho}\:,
\end{equation}
for $i=1,2,3,4$. Note that $d_1(t_0,r)$ is the only term in~\eqref{eq:dr_min}
that does not depend on parameter $\delta$.
The derivation of the DMT associated with the DoQF protocol will
be thus done as follows:
\begin{enumerate}
 \item Compute the terms $d_1(t_0,r)$, $d_2(t_0,\delta,r)$, $d_3(t_0,\delta,r)$
 and $d_4(t_0,\delta,r)$ for fixed values of $t_0$ and $\delta$ as given
 by~\eqref{eq:di_temp}. This is done in this Subsection.
 \item Compute $t_{0,\mathrm{DoQF}}^*(r)$ and $\delta_{0,\mathrm{DoQF}}^*(r)$ 
minimizing $d(t_0,\delta,r)$ defined from~\eqref{eq:dr_min} as the minimum of 
 $d_1(t_0,r)$, $d_2(t_0,\delta,r)$, $d_3(t_0,\delta,r)$
 and $d_4(t_0,\delta,r)$. This is done in Appendix \ref{app:proof}.
 \item The final DMT of the protocol can be finally obtained by calculating  
 $d(t_{0,\mathrm{DoQF}}^*(r),\delta_{0,\mathrm{DoQF}}^*(r),r)$.
This is done in Appendix \ref{app:proof}.
\end{enumerate}
 
\noindent {\bf Derivation of the term $\boldsymbol{d_1(t_0,r)}$, \textit{i.e.},
event $\boldsymbol{\mathcal{E}}$ is realized:}

Recall the definition given by~\eqref{eq:DF_po1} of $P_{o,1}(\rho)$ as the
probability that the destination is in outage and that the event $\mathcal{E}$
is realized. It is clear from~\eqref{eq:event_E} and~\eqref{eq:DF_po1} that
$P_{o,1}(\rho)$ is a function of parameter $t_0$. This is why the DMT term
$d_1(t_0,r)$ associated with $P_{o,1}(\rho)$ is also a function of this
parameter. Following the steps used in Appendix~\ref{app:der_d3}, one can
show that the following result holds.
\begin{equation}\label{eq:DF_d1}
d_1(t_0,r)=\left\{
\begin{array}{ll}
2(1-r)^+& \mbox{for}\: t_0\leq0.5\\
2-\frac{r}{1-t_0}& \mbox{for}\: t_0>0.5\: \mbox{and}\: r<1-t_0\\
\frac{(1-r)^+}{t_0}& \mbox{for}\: t_0>0.5\: \mbox{and}\: r\geq1-t_0
\end{array}
\right.
\end{equation}

\noindent {\bf Derivation of the term $\boldsymbol{d_2(t_0,\delta,r)}$,
\textit{i.e.}, events $\boldsymbol{\overline{\cal E}}$,
$\boldsymbol{\mathcal{S}}$ and $\boldsymbol{\mathcal{F}}$ are realized:}
 
Note from~\eqref{eq:event_F} and~\eqref{eq:DoQF_po3} that $P_{o,2}(\rho)$ is a
function of parameters $t_0$ and $\delta$. This is why the DMT
$d_2(t_0,\delta,r)$ associated with $P_{o,2}(\rho)$ is function of $t_0$ and
$\delta$.

First, consider the case $t_0\geq0.5$.

If parameter $\delta$ is chosen such that
$0<\delta\leq1-\left(1-\frac{r}{t_0}\right)^+$, then $d_2(t_0,\delta,r)$
can be written as
\begin{align}\label{eq:DoQF_d3}
&d_2(t_0,\delta,r)=\nonumber\\
&\left\{
\begin{array}{lll}
  (1-r)^++\max\left\{\left(1-\frac{r}{t_0}\right)^+,1-r-\delta\right\}, 
  &\frac{r}{t_1}-\left(1-\frac{r}{t_0}\right)^+-\frac{t_0}{t_1}\delta\leq 1-r\\ 
\frac{r}{t_1}-\left(1-\frac{r}{t_0}\right)^+-\frac{t_0}{t_1}\delta+
\max\left\{\frac{1-2r}{t_0}+\frac{t_1}{t_0}\left(1-\frac{r}{t_0}\right)^+,
\left(1-\frac{r}{t_0}\right)^+\right\}, 
  &\frac{r}{t_1}-\left(1-\frac{r}{t_0}\right)^+-\frac{t_0}{t_1}\delta> 1-r
\end{array}
\right.
\end{align}
As for the choice $\delta>1-\left(1-\frac{r}{t_0}\right)^+$, we show in
Appendix~\ref{app:der_d3} that event $\overline{\mathcal{E}}\&\mathcal{S}$
cannot be realized in this case for any channel state provided that $\rho$
is sufficiently large. Therefore, there exists $\rho_0>0$ such that 
$\forall \rho\geq\rho_0,\: P_{o,2}(\rho)=0$. The corresponding DMT
$d_2(t_0,\delta,r)$ will have no effect on the final DMT of the protocol.
The value $d_2(t_0,\delta,r)=2(1-r)^+$ is conveniently chosen in this case:
\begin{equation}\label{eq:DoQF_d3_triv}
d_2(t_0,\delta,r)=2(1-r)^+\: \textrm{ for}\:
\delta>1-\left(1-\frac{r}{t_0}\right)^+\:.
\end{equation}
The proof of~\eqref{eq:DoQF_d3} and~\eqref{eq:DoQF_d3_triv} is provided in
Appendix~\ref{app:der_d3}. We can show using the same arguments of the latter
appendix that 
\begin{equation}
d_2(t_0,\delta,r)=2(1-r)^+\:,\: \textrm{ for}\: \delta\leq 0\:.
\end{equation}
Similarly, we can obtain the expression \eqref{eq:DoQF_d3_bis}
for $d_2(t_0,\delta,r)$ in the case $t_0<0.5$.
\begin{align}\label{eq:DoQF_d3_bis}
&d_2(t_0,\delta,r)=\nonumber\\
&\left\{
\begin{array}{lll}
\left(1-\frac{r}{t_0}\right)^++\max\left\{(1-r)^+,\frac{1-r}{t_1}-
\frac{t_0}{t_1}\left(1-\frac{r}{t_0}\right)^+-\frac{t_0}{t_1}\delta\right\}, 
&\textrm{for}\: t_0<0.5\: \textrm{and}\: 2t_0 t_1\leq r\\ 
\left(1-\frac{r}{t_0}\right)^++\frac{r}{t_1}-\left(1-\frac{r}{t_0}
\right)^+-\frac {t_0}{t_1}\delta, &\textrm{for}\: t_0<0.5\: \textrm{and}\: 2t_0
t_1>r
\end{array}
\right.
\end{align}

\noindent {\bf Derivation of the term $\boldsymbol{d_3(t_0,\delta,r)}$, 
\textit{i.e.}, events $\boldsymbol{\overline{\cal E}}$,
$\boldsymbol{\mathcal{S}}$ and $\boldsymbol{\overline{\mathcal{F}}}$ are
realized:}

By referring to~\eqref{eq:event_F}
and~\eqref{eq:DoQF_po4}, it becomes clear that $P_{o,3}(\rho)$ 
is a function of parameters $t_0$ and $\delta$. This explains the fact that
$d_3(t_0,\delta,r)$ also depends on these two parameters.

The expression given below of $d_3(t_0,\delta,r)$ can be derived using the approach used 
in Appendix~\ref{app:der_d3}.
\begin{align}\label{eq:DoQF_d4}
d_3(t_0,\delta,r)=\left\{\begin{array}{ll}
2\left(1-\frac{r}{t_0}\right)^++\left(2\left(1-\frac{r}{t_0}\right)^++
\frac{t_0}{t_1}\delta-\frac{r}{t_1}\right)^+
&\mbox{for } \delta\leq1-\left(1-\frac{r}{t_0}\right)^+\\
2(1-r)^+ &\mbox{for } \delta>1-\left(1-\frac{r}{t_0}\right)^+
\end{array}\right.\:.
\end{align}
Recall that in the case $\delta>1-\left(1-\frac{r}{t_0}\right)^+$, event
$\overline{\mathcal{E}}\&\mathcal{S}$ cannot be realized, as we saw earlier, for
any channel realization provided that $\rho$ is sufficiently large. In this case
$P_{o,3}(\rho)=0$ and the corresponding DMT $d_3(t_0,\delta,r)$ will have no
effect on the final DMT of the protocol.
This is why the value $d_3(t_0,\delta,r)=2(1-r)^+$ was conveniently chosen
in~\eqref{eq:DoQF_d4} in this case.

\noindent {\bf Derivation of the term $\boldsymbol{d_4(t_0,\delta,r)}$, 
\textit{i.e.}, events $\boldsymbol{\overline{\cal E}}$ and
$\boldsymbol{\overline{\mathcal{S}}}$ are realized:}

This is the case when the relay does not quantize even if it has not succeeded
in decoding the source message. This happens when 
$\alpha_0\rho G_{01}+1<\Delta^2(\rho)$ which means that condition~\eqref{eq:Q_cond_b}
is not satisfied and the relay stays inactive.
Recall the definition of $P_{o,4}(\rho)$ as the probability that the destination
is in outage and that events $\overline{\mathcal{E}}$ and
$\overline{\mathcal{S}}$ are realized. It is straightforward to verify that
\begin{align}\label{eq:DoQF_d5}
d_4(t_0,\delta,r)=\left\{\begin{array}{ll}
(1-r)^++\max\left\{\left(1-\frac{r}{t_0}\right)^+,(1-\delta)^+\right\}
&\mbox{for } \delta>0\\
2(1-r)^+ &\mbox{for } \delta\leq0
\end{array}\right.\:.
\end{align}
Note that in the case $\delta\leq 0$, condition~\eqref{eq:Q_cond_b}
\emph{i.e.,} $\alpha_0\rho G_{01}+1>\Delta^2(\rho)$
is always satisfied for sufficiently large values of $\rho$ for all channel
realizations since $\Delta^2(\rho)\stackrel{.}{=}\rho^\delta\leq1$. 
Therefore, there exists in this case $\rho_0>0$ such that 
$\forall \rho\geq\rho_0$, event $\overline{\mathcal{S}}$ is never realized and
$P_{o,4}(\rho)=0$. The corresponding DMT $d_4(t_0,\delta,r)$ will have therefore
no effect on the final DMT of the protocol, and as usual 
we can assign it conveniently the value $d_4(t_0,\delta,r)=2(1-r)^+$ as done
in~\eqref{eq:DoQF_d5}.

\noindent {\bf Derivation of the final DMT of the DoQF protocol:}

At this point, the DMT terms $d_1(t_0,r)$, $d_2(t_0,\delta,r)$,
$d_3(t_0,\delta,r)$ and $d_4(t_0,\delta,r)$ associated with all the possible
cases encountered by the destination have been derived. the DMT
$d(t_0,\delta,r)$ associated with the DoQF protocol for
fixed values of $t_0$ and $\delta$ can now be obtained
from~\eqref{eq:DoQF_DMT_o} as the minimum of the above DMT terms.
No closed-form expression of $d(t_0,\delta,r)$ is given in this paper. However,
Theorem~\ref{the:DoQF_DMT} does provide the closed-form expression of 
$d_{\mathrm{DoQF}}^*(r)$ obtained by solving the optimization
problem $d_{\mathrm{DoQF}}^*(r)=
\sup_{\delta,t_0}d(t_0,\delta,r)$. The derivation of
$d_{\mathrm{DoQF}}^*(r)$ is provided in Appendix~\ref{app:proof}
and it leads to the expressions of $d_{\mathrm{DoQF}}^*(r)$,
$t_{0,\mathrm{DoQF}}^*(r)$ and
$\delta_{\mathrm{DoQF}}^*(r)$ given in
Theorem~\ref{the:DoQF_DMT}. 

\section{Numerical Illustrations and Simulations}
\label{sec:sim}

Simulations has been carried out assuming that channels are Rayleigh distributed
\emph{i.e.,} $H_{i,j}\sim\mathcal{CN}(0,\sigma_{i,j}^2)$. The corresponding
channel variance $\sigma_{i,j}^2$ is a function of the distance between
terminals following a path loss model with exponent equal to 3:
$\sigma_{i,j}^2=C d_{i,j}^{-3}$, where $d_{i,j}$ is the distance between nodes
$i$ and $j$, and the constant $C$ is chosen in such a way that
$\sigma_{0,2}^2=1$. The required data rate is equal
to 2 bits per channel use. 

\begin{figure}[h]
\centering\includegraphics[width=14cm]{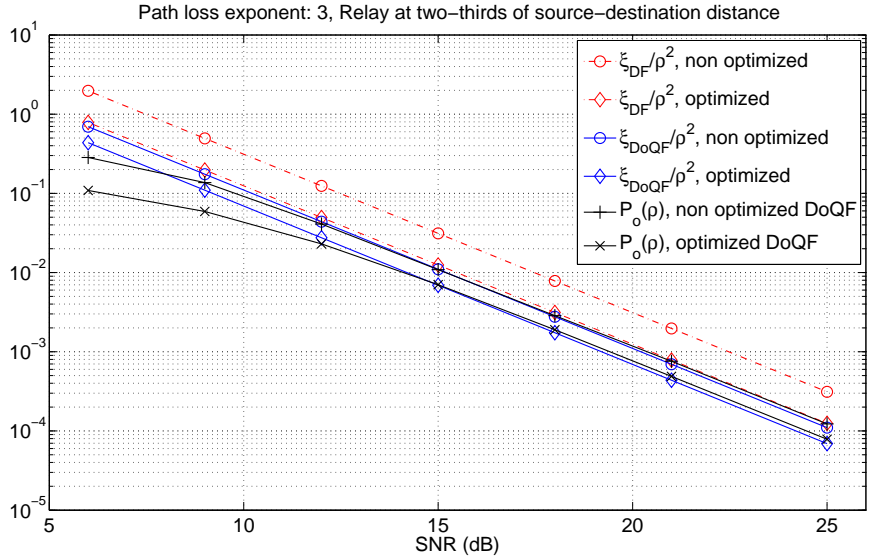}
\caption{Outage performance of the DF and DoQF protocols}
\label{fig:DoQF_DF}
\end{figure}

In Figure~\ref{fig:DoQF_DF}, outage probability performance with equal duration
time slots and equal amplitudes for both the DF and the DoQF (curves marked with
``non opt'') is compared to the performance after time and power optimization
(``opt'') for different values of the SNR $\rho$. Both the simulated outage
probability $P_o(\rho)$ and the approximated outage probability
$\frac{\xi_{DoQF}}{\rho^{2}}$ are plotted in this figure. The relay is assumed
to lie at two thirds of the source-destination distance on the
source-destination line segment. Substantial gains are observed between the DF
and the DoQF, and between optimized and non optimized protocols. Note that
minimizing the outage gain continues to reduce the outage probability of the
protocol even for moderate values of the SNR. 

\begin{figure}[h]
\centering\includegraphics[width=14cm]{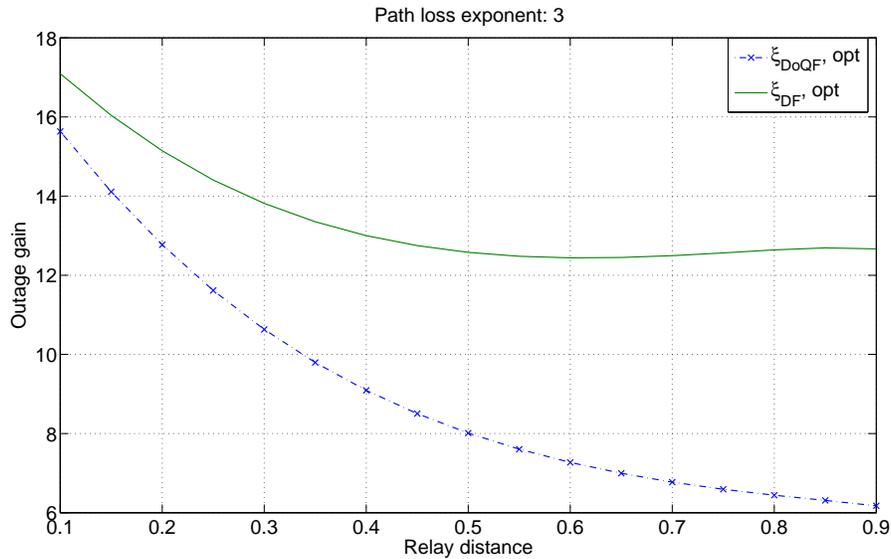}
\caption{Outage gain of DF and DoQF versus relay position}
\label{fig:out_dist}
\end{figure}

Figure~\ref{fig:out_dist} represents the outage gains for the DoQF and the DF
versus $d_{0,1}$, the position of the relay on the source-destination line
segment. Note from the figure that the farther the relay from the source is, the
better DoQF compared to DF works. This fact can be explained as follows: If the
relay is close to the destination, it will be more often in outage and the
Quantization step will thus operate more often.

\begin{figure}[h]
\centering\includegraphics[width=14cm]{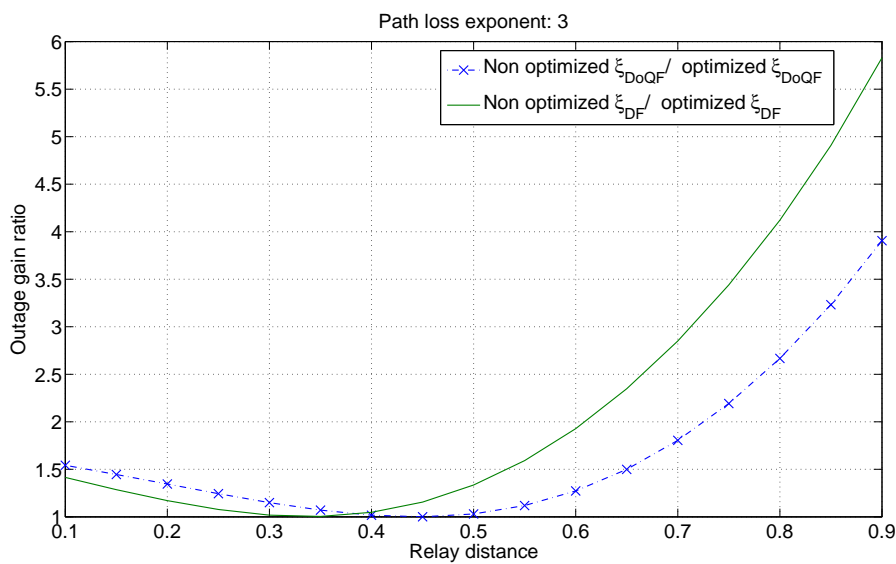}
\caption{Outage gain of DF and DoQF versus relay position}
\label{fig:out_opt}
\end{figure}

In Figure~\ref{fig:out_opt}, we plot the ratios of the outage gains with
equal times and equal powers to the optimized outage gains as a function of
the position $d_{0,1}$ of the relay on the source-destination segment. Note from
this figure that optimizing the slots durations and the power allocation yields
larger performance gains for both the DF and the DoQF when the relay is too
close or too far from the source.

\begin{figure}[h]
\centering\includegraphics[width=14cm]{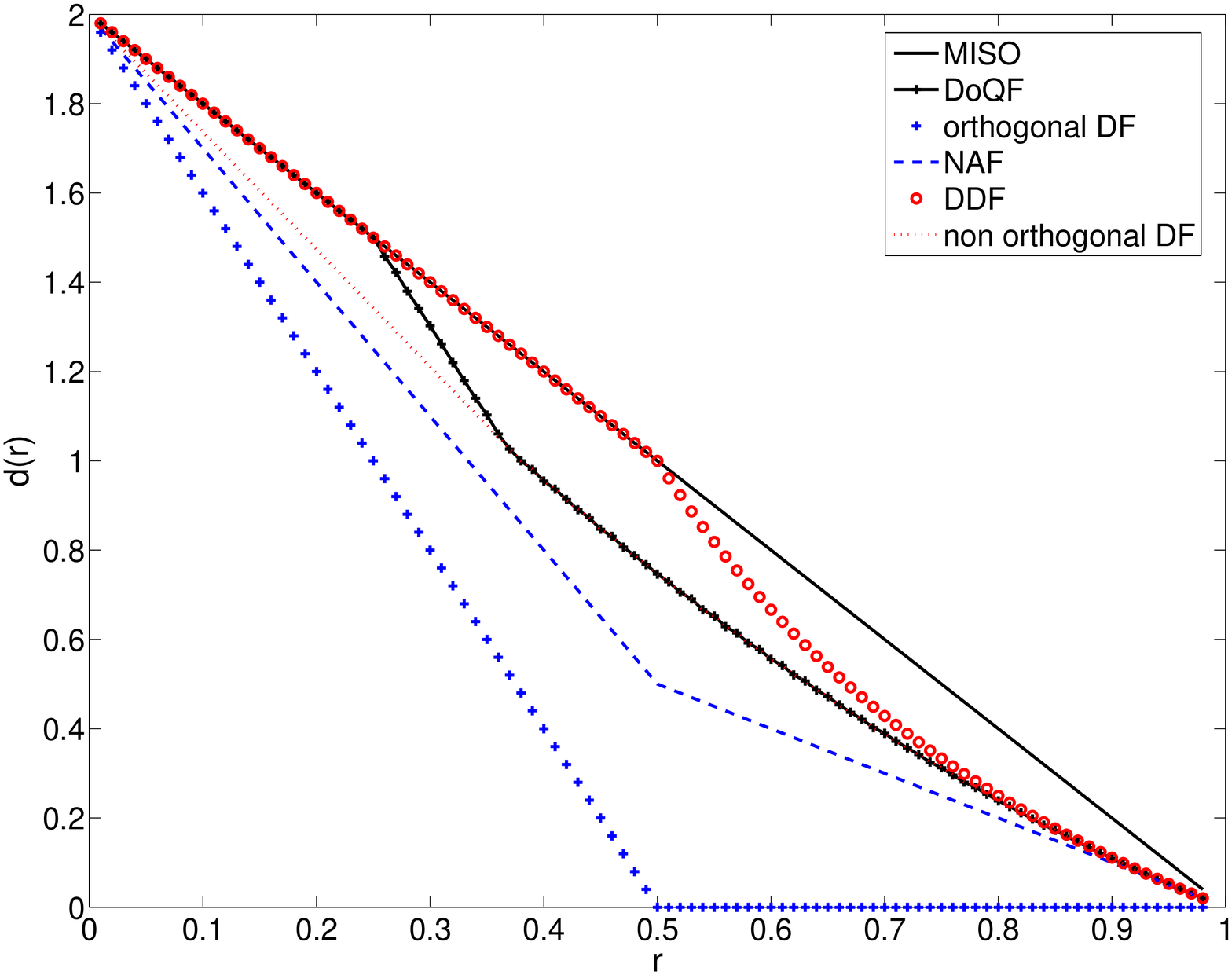}
\caption{DMT of the DF and DoQF protocols}
\label{fig:DMT}
\end{figure}

In Figure ~\ref{fig:DMT}, we plot the DMT of the DoQF (given by 
Theorem~\ref{the:DoQF_DMT}), orthogonal DF, (non-orthogonal) DF, NAF, DDF, and
the MISO bound. 

As already mentioned in a previous section, the DoQF outperforms the other
static protocols. In contrast, the DDF protocol is still better than the DoQF
but its dynamic approach leads to several implementation difficulties.  

\begin{figure}[h]
\centering\includegraphics[width=14cm]{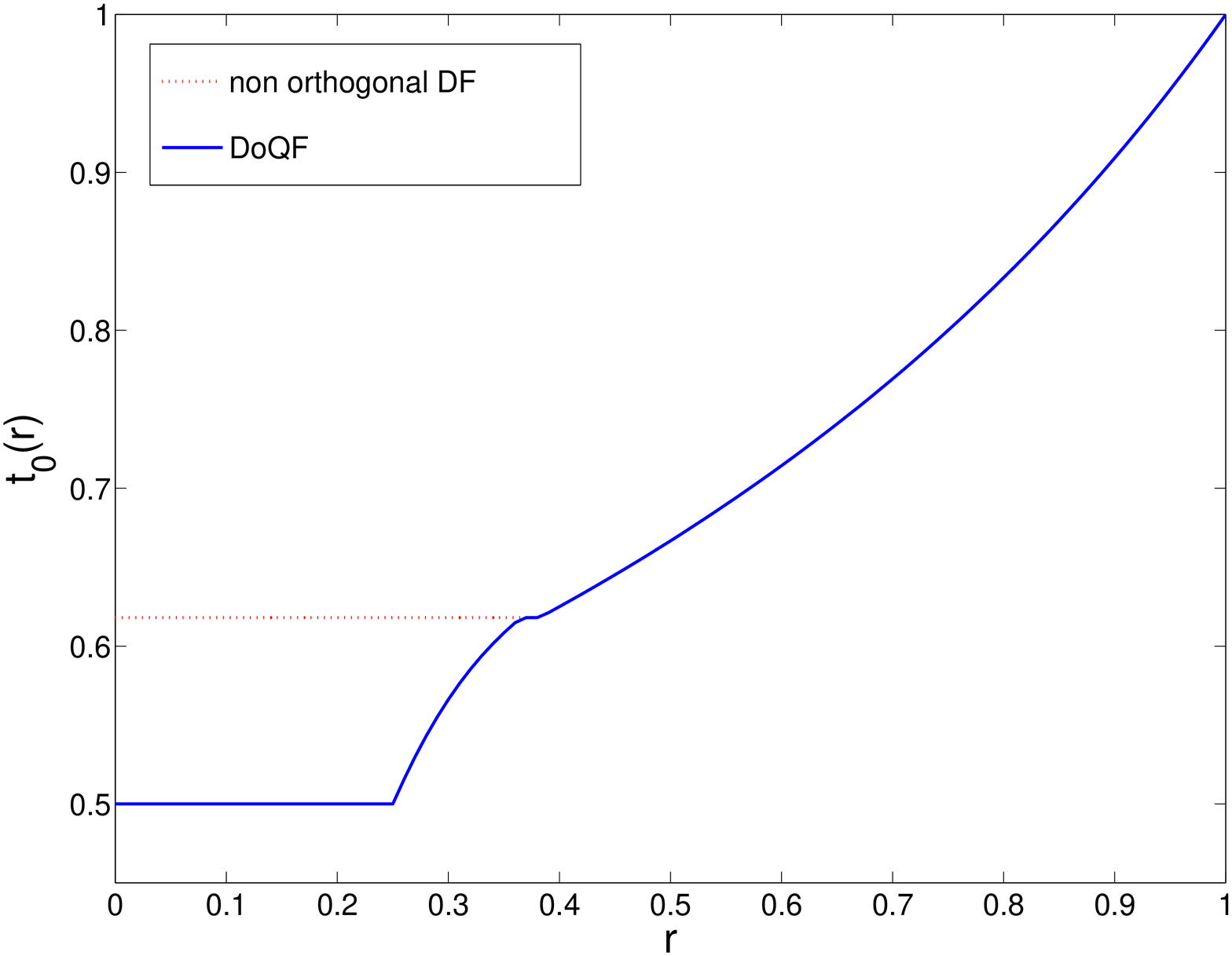}
\caption{Optimal $t_0$ for DF and DoQF}
\label{fig:t0r}
\end{figure}

In Figure \ref{fig:t0r}, the optimal sizes of slot 0 for the DoQF and the DF are
plotted. 
We remark that, when $r$ is small enough, slots 0 and 1 have the same length. 
When $r$ increases, the duration of relay listening increases also. As a consequence,
the duration of retransmission decreases. The duration for the quantization step thus 
decreases and the DoQF becomes closer to the DF as seen on the DMT.

\section{Conclusions}
\label{sec:concl}

A relaying protocol (DoQF) has been introduced for half-duplex single-relay
scenarios. The proposed DoQF is a static relaying protocol that involves
practical coding-decoding strategies at both the relay and the destination that
can be implemented in practice. The performance of this protocol has been
studied in the context of communications over slow fading wireless channels
using two relevant performance metrics: The outage gain and the diversity
multiplexing tradeoff (DMT).
The DoQF protocol has been shown to be optimal in terms of outage gain in the
wide class of half-duplex static relaying protocols. A method to minimize the
outage gain of the DoQF w.r.t the slots durations and the power allocation has
been also proposed.
The proposed protocol has been finally shown to achieve the DMT of MISO for
multiplexing gains $r\leq 0.25$.
Some future research directions would be the extension of the proposed DoQF
protocol to multi-relay networks and to networks involving frequency selective
channels.

\appendices
\section{Proof of Theorem~\ref{the:genie}}
\label{app:genie_proof}

It is known~\cite{coop} that the capacity of any static relaying protocol 
is limited by the cut-set upper-bound. In this appendix, we derive the outage
gain associated with the cut-set capacity. We prove next that this
outage gain is equal to $\xi_{\textrm{CS-HD}}$ given by~\eqref{eq:genie}.
 
The cut-set upper-bound on the capacity of any half-duplex single-relay
protocol from the class $\mathcal{P}_{\textrm{HD}}(t_0,\alpha_0,\alpha_1)$, with
a listening time equal to $t_0 T$ and a cooperation time equal to 
$(1-t_0)T=t_1 T$, is given by
\begin{eqnarray}
C_{\textrm{CS-HD}}=\lim_{T\to\infty}\frac{1}{T}\max_{p(X_{00},X_{01},X_{11})}
\min\big\{
I(X_{00};Y_{10},Y_{20})+I(X_{01};Y_{21}|X_{11}),\nonumber\\
I(X_{00};Y_{20})+I(X_{01},X_{11};Y_{21})\big\}\:,
\label{eq:cut_set}
\end{eqnarray}
where the maximization in~\eqref{eq:cut_set} is with respect to all the joint
distributions of $X_{00}$, $X_{01}$ and $X_{11}$ that satisfy the power
constraints~\eqref{eq:high_snr_constraint0} and~\eqref{eq:high_snr_constraint1}.
It can be shown that the maximum in~\eqref{eq:cut_set} is achieved when vectors
$X_{00}$, $X_{01}$ and $X_{11}$ are zero-mean i.i.d Gaussian with covariance
matrices that satisfy constraints~\eqref{eq:high_snr_constraint0}
and~\eqref{eq:high_snr_constraint1}. The cut-set upper-bound can thus be written
as
\begin{align}
C_{\textrm{CS-HD}}=\min\Big\{&
t_0\log\left(1+\mathbb{E}\left[|X_0(i)|^2\right]G_{01}+
\mathbb{E}\left[|X_0(i)|^2\right] G_{02}\right)+
t_1\log\left(1+\mathbb{E}\left[|X_0(i)|^2\right]G_{02}\right),\nonumber\\
&t_0\log\left(1+\mathbb{E}\left[|X_0(i)|^2\right]G_{02}\right)+
t_1\log\left(1+\mathbb{E}\left[|X_0(i)|^2\right]G_{02}+
\mathbb{E}\left[|X_{11}(i)|^2\right]G_{12}\right)\Big\}\nonumber\\
=\min\{&C_{\textrm{SIMO}},C_{\textrm{MISO}}\}\:,
\label{eq:cut_set_gaussian}
\end{align}
where $C_{\textrm{SIMO}}$ and $C_{\textrm{MISO}}$ are defined in order to
simplify the presentation of the proof as follows:
\begin{align*}
C_{\textrm{SIMO}}&=t_0\log\left(1+\mathbb{E}\left[|X_0(i)|^2\right]G_{01}+
\mathbb{E}\left[|X_0(i)|^2\right]G_{02}\right)+
t_1\log\left(1+\mathbb{E}\left[|X_0(i)|^2\right]G_{02}\right)\\
C_{\textrm{MISO}}&=t_0\log\left(1+\mathbb{E}\left[|X_0(i)|^2\right]G_{02}
\right)+
t_1\log\left(1+\mathbb{E}\left[|X_0(i)|^2\right]G_{02}+
\mathbb{E}\left[|X_{11}(i)|^2\right]G_{12}\right)\:.
\end{align*}
We now prove that the limit 
$\lim_{\rho\to\infty}\rho^2\mathrm{Pr}[C_{\mathrm{CS-HD}}\leq R]$ exists and
that it is equal to $\xi_{\mathrm{CS-HD}}$ given by~\eqref{eq:genie}. For that
sake, note that the following holds:
\begin{align*}
\mathrm{Pr}[C_{\textrm{CS-HD}}\leq R]=
&1-\mathrm{Pr}\left[C_{\textrm{CS-HD}}>R\right]\\
=&1-\mathrm{Pr}[C_{\textrm{SIMO}}>R, C_{\textrm{MISO}}>R]\\
\geq&1-\mathrm{Pr}\left[C_{\textrm{SIMO}}>R\right]\times
\mathrm{Pr}\left[C_{\textrm{MISO}}>R\right]\\
=&1-\left(1-\mathrm{Pr}\left[C_{\textrm{SIMO}}\leq R\right]\right)
\times\left(1-\mathrm{Pr}\left[C_{\textrm{MISO}}\leq R\right]\right)\:.
\end{align*}
Now define
\begin{align*}
P_{o,\textrm{SIMO}}&=\mathrm{Pr}\left[
C_{\textrm{SIMO}}\leq R\right]\\
P_{o,\textrm{MISO}}&=\mathrm{Pr}[C_{\textrm{MISO}}\leq R]\:.
\end{align*}
Using these new notations, we conclude that the following lower-bound on
$\mathrm{Pr}[C_{\textrm{CS-HD}}\leq R]$ holds:
\begin{equation}\label{eq:simo_miso}
\mathrm{Pr}[C_{\textrm{CS-HD}}\leq R]\geq
P_{o,\textrm{SIMO}}+P_{o,\textrm{MISO}}
-P_{o,\textrm{SIMO}} P_{o,\textrm{MISO}}\:.
\end{equation}
In the same way, it is straightforward to show that
$\mathrm{Pr}[C_{\textrm{CS-HD}}\leq R]$ can be upper-bounded as follows.
\begin{equation}\label{eq:simo_miso_upper}
\mathrm{Pr}[C_{\textrm{CS-HD}}\leq R]\leq
P_{o,\textrm{SIMO}}+P_{o,\textrm{MISO}}
+P_{o,\textrm{SIMO}} P_{o,\textrm{MISO}}\:.
\end{equation}
Now, we can use the same arguments and tools employed in
Subsection~\ref{sec:DoQF_out} to prove that
\begin{align}
 &\lim_{\rho\to\infty}\rho^2
P_{o,\textrm{SIMO}}=\frac{c_{02}c_{01}}{\alpha_0^2}\int_{\mathbb{R }_+^2}
 \mathbf{1}\left\{t_1\log(1+u)+t_0\log\left(1+u+v\right)\leq R\right\}
 du dv\label{eq:po_simo}\\
 &\lim_{\rho\to\infty}\rho^2
P_{o,\textrm{MISO}}=\frac{c_{02}c_{12}}{\alpha_0\alpha_1}\int_{\mathbb{R}_+^2}
\mathbf{1}\{t_0\log(1+u)+t_1\log(1+u+v)\leq R\}du dv\label{eq:po_miso}\\
 &\lim_{\rho\to\infty} \rho^2 P_{o,\textrm{SIMO}}
P_{o,\textrm{MISO}}=0\:.\label{eq:product}
\end{align}
Note that the integrals in the rhs of~\eqref{eq:po_simo} and~\eqref{eq:po_miso}
coincide with the two integrals in the rhs of~\eqref{eq:DoQF_po_lim}. We can
thus write 
\begin{align}
 \lim_{\rho\to\infty}\rho^2 P_{o,\textrm{SIMO}}&=
 \frac{c_{02}c_{01}}{\alpha_0^2}\left(\frac{1}{2}+
 \frac{\exp(2R)}{4t_0-2}-\frac{t_0\exp(R/t_0)}{2t_0-1}\right)\label{eq:simo}\\
 \lim_{\rho\to\infty}\rho^2 P_{o,\textrm{MISO}}&=\frac{c_{02}c_{12}}
 {\alpha_{01}\alpha_{02}}\left(\frac{1}{2}+\frac{\exp(2R)}{4t_1-2}-
\frac{t_1\exp(R/t_1)}{ 2t_1-1} \right)\:.\label{eq:miso}
\end{align}

Combining~\eqref{eq:simo_miso}, \eqref{eq:simo_miso_upper}, \eqref{eq:product},
\eqref{eq:simo}
and~\eqref{eq:miso} we conclude that
\begin{align*}
\lim_{\rho\to\infty}\rho^2 \mathrm{Pr}[C_{\textrm{CS-HD}}\leq RT]=
\xi_{\mathrm{CS-HD}}\:,
\end{align*}
where $\xi_{\textrm{CS-HD}}$ is the lower-bound defined by~\eqref{eq:genie}. 
Note that since $C_{\textrm{CS-HD}}$ is an upper-bound on the capacity of any
static half-duplex relaying protocol belonging to the class
$\mathcal{P}_{\textrm{HD}}(t_0,\alpha_0,\alpha_1)$, then $\xi_{\textrm{CS-HD}}$
which satisfies $\xi_{CS-HD}=$ $\lim_{\rho\to\infty}$
$\rho^2\mathrm{Pr}[C_{\textrm{CS-HD}}\leq RT]$
is a lower-bound on the outage gain of any protocol from the class
$\mathcal{P}_{\textrm{HD}}(t_0,\alpha_0,\alpha_1)$.
This completes the proof of Theorem~\ref{the:genie}.

\section{Derivation of $d_2(t_0,\delta,r)$ (for $t_0\geq0.5$
and $\delta>0$)}
\label{app:der_d3}

First, recall the definition of $d_2(t_0,\delta,r)$ as
$d_2(t_0,\delta,r)=-\lim_{\rho\to\infty}\frac{\log(P_{o,2}(\rho))}{\log\rho}$,
where the probability $P_{o,2}(\rho)$ is defined by~\eqref{eq:DoQF_po3} as
\begin{align}
P_{o,2}(\rho)=\mathrm{Pr}\Bigg[&t_1 \log(1+\alpha_0\rho
G_{02})+t_0\log\left(1+\alpha_0\rho
G_{02}+\frac{\gamma(G_{01},\rho)\alpha_0\rho G_{01}}
{\gamma(G_{01},\rho)+\Delta^2(\rho)\sqrt{\gamma(G_{01},\rho))}}\right)
\leq R(\rho),\nonumber\\
&\overline{\mathcal{E}}, \mathcal{F},
\mathcal{S}\Bigg]\:,\label{eq:DoQF_po3_bis}
\end{align}
where $\gamma(G_{01},\rho)=\frac{
\left(1+\alpha_0\rho G_{01}-\Delta^2(\rho)\right)^2}
{\left(1+\alpha_0\rho G_{01}\right)^2}$, and where events~$\mathcal{E}$,
$\mathcal{S}$ and $\mathcal{F}$ are defined by~\eqref{eq:event_E},
\eqref{eq:event_S} and~\eqref{eq:event_F} respectively. Note that 
$\gamma(G_{01},\rho)$ is positive since event 
$\mathcal{S}$ \emph{i.e.,} $1+\alpha_0\rho G_{01}\geq\Delta^2(\rho)$, is
realized. Furthermore, we can check that the following result holds.
\begin{equation}\label{eq:gamma_rho}
\frac{\gamma(G_{01},\rho)}
{\gamma(G_{01},\rho)+\Delta^2(\rho)\sqrt{\gamma(G_{01},\rho)}}\stackrel{.}{=}
\frac{1}{1+\Delta^2(\rho)}\stackrel{.}{=}\rho^{-(\delta)^+}\:.
\end{equation}

In the following, we assume that $R(\rho)=r\log\rho$ in accordance
with~\eqref{eq:DMT_def}, and we define as in~\cite{tse} the \emph{exponential
order} associated with channel $H_{ij}$ as 
$a_{ij}=-\frac{\log G_{ij}}{\log\rho}$.
We can easily verify that $a_{ij}$ is a \emph{Gumbel} distributed random
variable with the probability density function 
$f_{a_{ij}}(a)=\log\rho\: e^a e^{-e^{-a\log\rho}}$.
By plugging $G_{01}=\rho^{-a_{01}}$ into~\eqref{eq:event_E}, the probability of
the event $\overline{\mathcal{E}}$ \emph{i.e.,} 
$t_0\log(1+\alpha_0\rho G_{01})>R(\rho)$, can be written as
\begin{equation}\label{eq:event_E_bis}
\mathrm{Pr}[\mathcal{E}]\dot{=}\mathrm{Pr}\left[\left(1-a_{01}\right)^+\leq\frac
{r}{t_0}\right]\:.
\end{equation}
Similarly, we can verify that the probability of event $\mathcal{F}$ 
\emph{i.e.,} $t_1\log\left(1+\frac{\phi(\rho)G_{12}}{\alpha_0\rho
G_{02}+1}\right)>Q(\rho)t_0$, satisfies
\begin{equation}\label{eq:event_F_bis}
\mathrm{Pr}[\mathcal{F}]\dot{=}\mathrm{Pr}\left[\left(1+\left(1-\frac{r}{t_0}
\right)^+-a_{12}-(1-a_{02})^+\right)^+
\leq\frac{r}{t_1}-\frac{t_0}{t_1}\delta\right]\:,
\end{equation}
and that the probability of $\mathcal{S}$ satisfies
\begin{equation}\label{eq:event_S_bis}
\mathrm{Pr}[\mathcal{S}]\dot{=}\mathrm{Pr}[\delta\leq\left(1-a_{01}\right)^+]\:.
\end{equation}
By plugging $R(\rho)=r\log\rho$, $G_{01}=\rho^{-a_{01}}$,
$G_{02}=\rho^{-a_{02}}$,
$G_{12}=\rho^{-a_{12}}$, \eqref{eq:gamma_rho}, \eqref{eq:event_E_bis},
\eqref{eq:event_F_bis} and~\eqref{eq:event_S_bis} into~\eqref{eq:DoQF_po3_bis},
the following high SNR result holds for $\delta>0$.
\begin{align}
P_{o,2} (\rho) \dot{=}
\mathrm{Pr}\Bigg[&t_1(1-a_{02})^++t_0(1-\min(a_{02},a_{01}+\delta))^+<r\:,	
\:(1-a_{01})^+ < \frac{r}{t_0}\:,\nonumber\\					
&\:\left(1+\left(1-\frac{r}{t_0}\right)^+-a_{12}-(1-a_{02})^+\right)^+>\frac{r}{
t_1}-\frac{t_0}{t_1}\delta\:,
\:\delta\leq\left(1-a_{01}\right)^+\Bigg],
\end{align}
or, equivalently, 
\begin{equation}\label{eq:po3_o}
P_{o,2}(\rho)\dot{=}\int_{\mathcal{O}}f_{a_{01}}(a_{01})f_{a_{02}}(a_{02})f_{a_{
12}}
(a_{12})da_{01}da_{02}da_{12}\:,
\end{equation}
where $f_{a_{ij}}(.)$ is the probability density function of $a_{ij}$ and 
\begin{align}\label{eq:o}
\mathcal{O}=\Bigg\{
&(a_{01},a_{02},a_{12})\in\mathbb{R}^3\:|\:t_1(1-a_{02}
)^++t_0(1-\min(a_{02},a_{01}+\delta))^+<r\:,
\:(1-a_{01})^+ < \frac{r}{t_0}\:,\nonumber\\
&\:\left(1+\left(1-\frac{r}{t_0}\right)^+-a_{12}-(1-a_{02})^+\right)^+
>\frac{r}{t_1}-\frac{t_0}{t_1}\delta\:,\:\delta\leq\left(1-a_{01}
\right)^+
\Bigg\}\:.
\end{align}
Plugging the expression of $f_{a_{ij}}(.)$ given earlier into~\eqref{eq:po3_o},
$P_{o,2}(\rho)$ can be written as
$$P_{o,2}(\rho)\dot{=}\int_{\cal O}(\log\rho)^3
\rho^{-(a_{01}+a_{02}+a_{12})}e^{-\rho^{-a_{01}}}e^{-\rho^{-a_{02}}}e^{-\rho^{
-a_{12}}}
da_{01}da_{02}da_{12}\:.$$
It can be shown (refer to~\cite{tse}) that the term $(\log\rho)^3$ can be
dropped from the latter equation without losing its exactness. Moreover,
integration in the same equation can be restricted to positive values
of $a_{01}$, $a_{02}$ and $a_{12}$. Define
$\mathcal{O}_+=\mathcal{O}\cap\mathbb{R}_+^3$. The probability
$P_{o,2}(\rho)$ thus satisfies
\begin{equation}\label{eq:po3_o+}
P_{o,2}(\rho)\dot{=}\int_{\mathcal{O}_+}\rho^{-(a_{01}+a_{02}+a_{12})}
da_{01}da_{02}da_{12}\:,
\end{equation}
and the DMT $d_2(t_0,\delta,r)$ associated with $P_{o,2}(\rho)$ can now be
written~\cite{tse} as
\begin{equation}\label{eq:infimum}
d_2(t_0,\delta,r)=\inf_{(a_{01},a_{02},a_{12})\in{{\cal
O}^+}}(a_{01}+a_{02}+a_{12})\:.
\end{equation}
In this appendix, the derivation of $d_2(t_0,\delta,r)$ will be done only in the
case characterized by $t_0\geq0.5$ and $\delta>0$. The derivation in the case
$\delta\leq0$ or $t_0<0.5$ follows the same approach. 

\begin{figure}[h]
\begin{minipage}[b]{0.49\linewidth}
\centering
\includegraphics[scale=0.50]{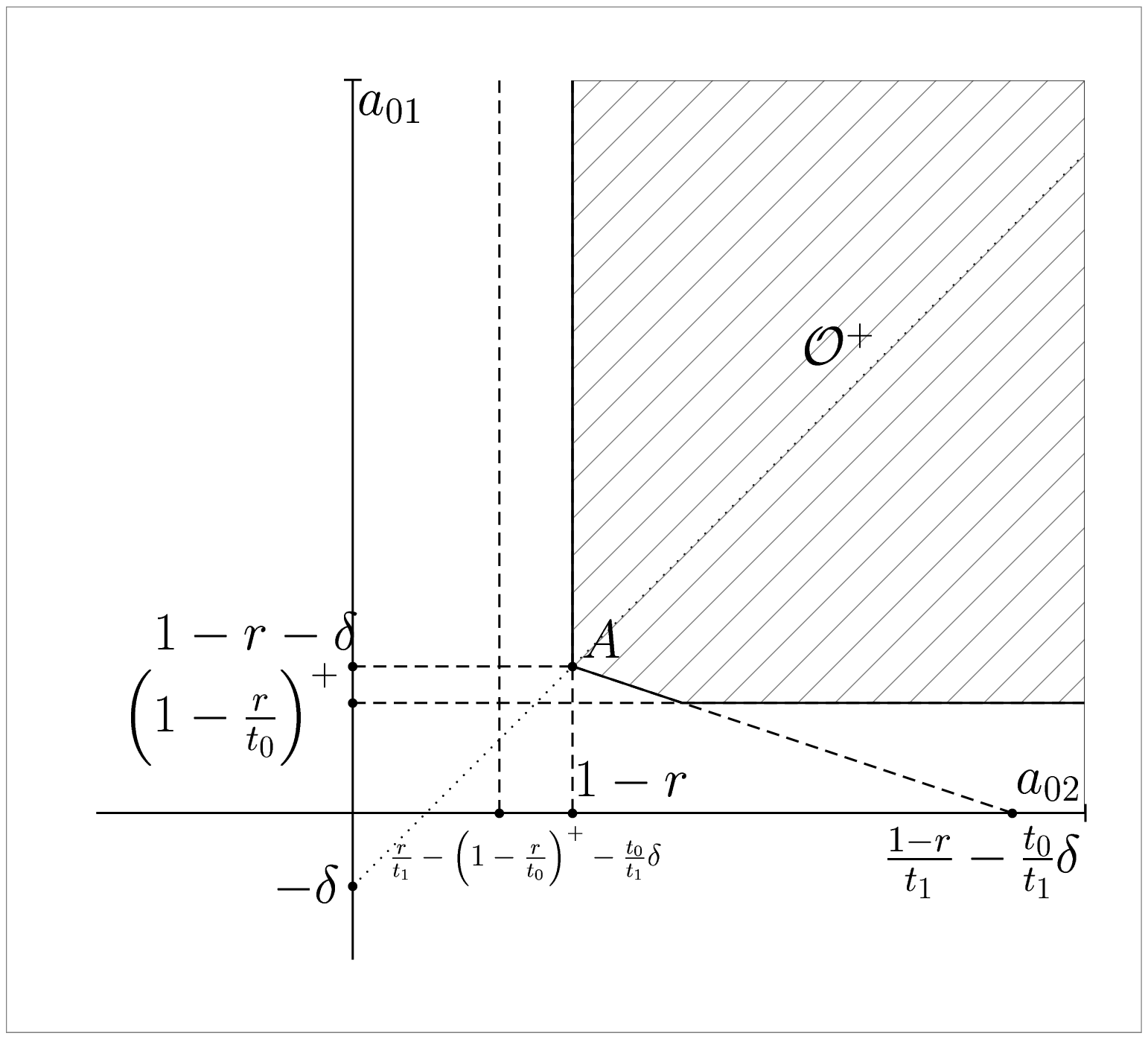}
\caption{Outage region for the DoQF protocol in the case
$\frac{r}{t_1}-\left(1-\frac{r}{t_0}\right)^+$ $-\frac{t_0}{t_1}\delta\leq
1-r$.}
\label{fig:DoQF_O+2}
\end{minipage}\hfill
\begin{minipage}[b]{0.49\linewidth}
\centering
\includegraphics[scale=0.50]{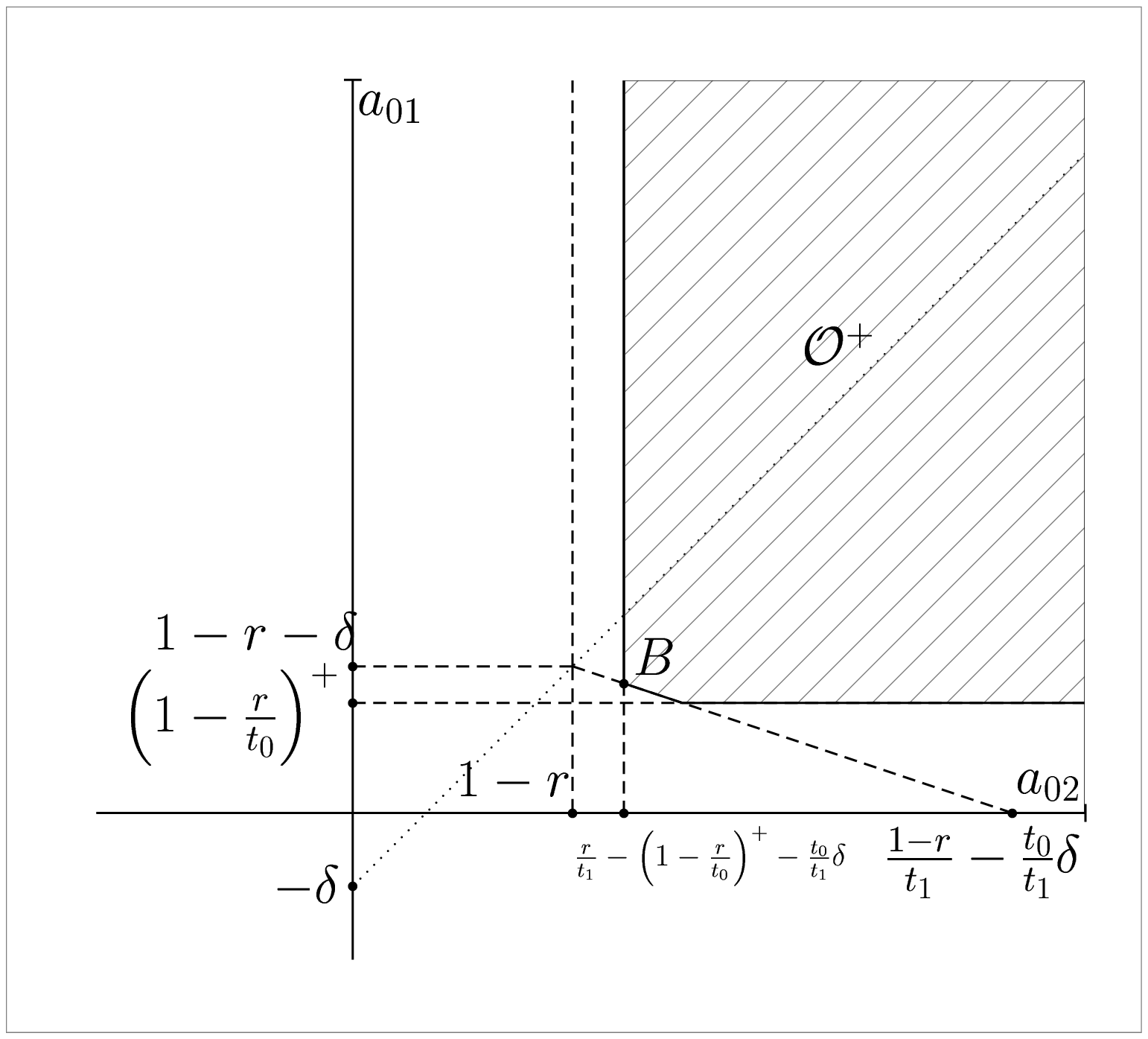}
\caption{Outage region for the DoQF protocol in the case
$1-r<\frac{r}{t_1}-\left(1-\frac{r}{t_0}\right)^+-\frac{t_0}{t_1}\delta$.}
\label{fig:DoQF_O+3}
\end{minipage}
\end{figure}

Consider first the case $0<\delta\leq1-\left(1-\frac{r}{t_0}\right)^+$.
The infimum in~\eqref{eq:infimum} can be computed by partitioning ${\cal O}_+$
into subsets according to whether $a_{01}$, $a_{02}$ are smaller or larger
than 1.
\begin{itemize}
\item[$\bullet$] $\boldsymbol{a_{01}>1}$. 
In this case, $(1-a_{01})^+=0$ and the fourth inequality in~\eqref{eq:o}
reduces to $\delta\leq0$. This result contradicts our assumption that
$\delta>0$. There is therefore no triples 
$(a_{01},a_{02},a_{12})\in{\cal O}^+$ such that $a_{01}>1$.
\item[$\bullet$] $\boldsymbol{a_{01}\leq1, a_{02}>1}$. Since the third
inequality in the definition of $\cal O$ given by~\eqref{eq:o} contains the term
$\left(1+\left(1-\frac{r}{t_0}\right)^+-a_{12}-(1-a_{02})^+\right)^+$, then we
should consider two categories of triples $(a_{01},a_{02},a_{12})$:
\begin{itemize}
 \item[$\circ$] $1+\left(1-\frac{r}{t_0}\right)^+-a_{12}-(1-a_{02})^+<0$.\\ 
 For triples $(a_{01},a_{02},a_{12})\in{\cal O}^+$ under this category, the
 third inequality in~\eqref{eq:o} can be reduced to $\delta>\frac{r}{t_0}$,
 which contradicts the second and the fourth inequalities in~\eqref{eq:o}. This
 category can be therefore dropped out.
 \item[$\circ$] $1+\left(1-\frac{r}{t_0}\right)^+-a_{12}-(1-a_{02})^+\geq0$.\\ 
 Recall the first inequality in~\eqref{eq:o} \emph{i.e.,}
 $t_1(1-a_{02})^++t_0(1-\min(a_{02},a_{01}+\delta))^+<r$.
 Since $\delta\leq(1-a_{01})^+$ due to the fourth inequality in~\eqref{eq:o},
 then $a_{01}+\delta\leq a_{01}+(1-a_{01})^+=1
 \leq a_{02}$. The first inequality in~\eqref{eq:o} reduces thus to 
 $a_{01}\geq\left(1-\frac{r}{t_0}\right)^+$. We conclude that
 \begin{equation}\label{eq:inf1}
  \inf_{a_{01}\leq 1, a_{02}>1}(a_{01}+a_{02}+a_{12})=
  1+\left(1-\frac{r}{t_0}\right)^+\:.
 \end{equation}
 One can verify after some simple algebra that $\inf_{a_{01}\leq
 1,a_{02}>1}(a_{01}+a_{02}+a_{12})=
 1+\left(1-\frac{r}{t_0}\right)^+$ is always larger than $d_1(t_0,r)$ given
 by~\eqref{eq:DF_d1}. Therefore, the term $\inf_{a_{01}\leq 1,
 a_{02}>1}(a_{01}+a_{02}+a_{12})$ never coincides with the minimum in
 $d(t_0,\delta,r)=\min\{d_1(t_0,r)$, $d_2(t_0,\delta,r)$,$d_3(t_0,\delta,r)$,
 $d_4(t_0,\delta,r)\}$. As a result, the argument of the infimum
 $\inf_{(a_{01},a_{02},a_{12})\in{{\cal O}_+}}(a_{01}+a_{02}+a_{12})$ coincides
 necessarily with a triple $(a_{01},a_{02},a_{12})$ from the following subset.
\end{itemize}
\item[$\bullet$] $\boldsymbol{a_{01}\leq1, a_{02}\leq1}$. Two categories of
triples $(a_{01},a_{02},a_{12})$ should be considered.
\begin{itemize}
 \item[$\circ$] $1+\left(1-\frac{r}{t_0}\right)^+-a_{12}-(1-a_{02})^+<0$.\\
 As done before, it is straightforward to verify that there is no triples
 $(a_{01},a_{02},a_{12})\in{\cal O}^+$ that fall under this category.
 \item[$\circ$] $1+\left(1-\frac{r}{t_0}\right)^+-a_{12}-(1-a_{02})^+\geq0$.\\
 The third inequality in~\eqref{eq:o} leads in this case to
 \begin{equation}\label{eq:a02}
 a_{02}>\frac{r}{t_1}-\left(1-\frac{r}{t_0}\right)^+-\frac{t_0}{t_1}\delta\:.
 \end{equation}
 In order to evaluate the first inequality in~\eqref{eq:o}, two subcategories
 of triples $(a_{01},a_{02},a_{12})$ should be further examined.
 \begin{enumerate}
  \item $a_{02}<a_{01}+\delta$. For triples 
  $(a_{01},a_{02},a_{12})\in{\cal O}^+$ under this category, the first
  inequality in~\eqref{eq:o} leads to $a_{02}>(1-r)^+$.
  \item $a_{02}\geq a_{01}+\delta$. The first inequality results in this case in
  $a_{02}+\frac{t_0}{t_1}a_{01}>\frac{1-r}{t_1}-\frac{t_0}{t_1}\delta$.
 \end{enumerate}
Referring to Figures~\ref{fig:DoQF_O+2} and~\ref{fig:DoQF_O+3} reveals that
$\inf_{a_{01}\leq 1,a_{02}\leq1}(a_{01}+a_{02}+a_{12})$ coincides with the rhs
of~\eqref{eq:DoQF_d3}. We have thus proved that $d_2(t_0,\delta,r)$ is indeed
given by~\eqref{eq:DoQF_d3}.
\end{itemize}
\end{itemize}
Now consider the case $\delta>1-\left(1-\frac{r}{t_0}\right)^+$ in order to
prove that~\eqref{eq:DoQF_d3_triv} holds.
To that end, refer to the second and the fourth inequalities in the definition
of $\mathcal{O}$ given by~\eqref{eq:o}, that is
$(1-a_{01})^+ < \frac{r}{t_0}$ and $\delta\leq(1-a_{01})^+$. Note that
$(1-a_{01})^+\leq1$ since $a_{01}>0$. A necessary condition for $a_{01}$
to satisfy the second and the fourth inequalities in~\eqref{eq:o}, and
consequently to belong to $\mathcal{O}_+$ is thus
$\delta\leq\min\left\{1,\frac{r}{t_0}\right\}=1-\left(1-\frac{r}{t_0}\right)^+$.
This means that if we choose $\delta$ such that
$\delta>1-\left(1-\frac{r}{t_0}\right)^+$, the set $\mathcal{O}_+$ will be
empty. In this case, $P_{o,2}(\rho)=0$ for sufficiently large $\rho$. In other
words, there exists $\rho_0>0$ such that $\forall \rho\geq \rho_0$, the event
$\overline{\mathcal{E}}\&\mathcal{S}$ cannot be realized and the relay will not
be able to quantize, reducing the DoQF to a classical DF scheme. The
corresponding DMT $d_2(t_0,\delta,r)$ will have no effect in this case on the
final DMT of the protocol. We can give it for convenience the value
$d_2(t_0,\delta,r)=2(1-r)^+$, which is the upper-bound on the DMT of any
single-relay protocol.

\section{Derivation of $d_{\mathrm{DoQF}}^*(r)=
\sup_{\delta,t_0}d(t_0,\delta,r)$}
\label{app:proof}

Before proceeding with the proof, it is useful to recall here the definition of
$t_{0,{\mathrm{DoQF}}}^*(r)$ and $\delta_{\mathrm{DoQF}}^*(r)$ as the argument
of the supremum in $d_{\mathrm{DoQF}}^*(r)=\sup_{\delta, t_0}d(t_0,\delta,r)$. 

We will first compute $d_{\mathrm{DoQF}}^*(r)$ in the case $r\leq0.25$, and then
in the case $r>0.25$.

\noindent {\bf The case $\boldsymbol{r\leq0.25}$}

Let us plug $t_0=0.5$ and $\delta=0$
into~\eqref{eq:DF_d1}, \eqref{eq:DoQF_d3}, \eqref{eq:DoQF_d4}
and~\eqref{eq:DoQF_d5} to obtain
\begin{align}
 &d_1(t_0,r)=d_2(t_0,\delta,r)=d_4(t_0,\delta,r)=2(1-r)^+\:,\\
 &d_3(t_0,\delta,r)=2(1-2r)^++\left(2(1-2r)^+-2r\right)^+=2-8r\:.
 \label{eq:d4_miso}
\end{align}
Note that $d_3(t_0,\delta,r)$ is the only term that may be different from
$2(1-r)^+$. However, one can verify by referring to~\eqref{eq:d4_miso} that
$d_3(t_0,\delta,r)\geq2(1-r)^+\:\Leftrightarrow\:r\leq0.25$. We conclude that,
for $r\leq0.25$, $d(0.5,0,r)=2(1-r)^+$. We have thus proved that the MISO
upper-bound is achieved by the DoQF for $r\leq 0.25$ by choosing
$t_{0,\textrm{DoQF}}^*(r)=0.5$ and $\delta_{\textrm{DoQF}}^*(r)=0$. 

\noindent
{\bf The case $\boldsymbol{r>0.25}$}

The first step of the proof in this case is to reduce the size of the set of
possible values of $t_{0,{\mathrm{DoQF}}}^*(r)$ and
$\delta_{\mathrm{DoQF}}^*(r)$. We will prove in particular that
the following three lemmas hold.
\begin{lemma}\label{lem:DoQF_DF}
For any $r\in[0,1]$, $d_{\mathrm{DoQF}}^*(r)\geq d_{\mathrm{DF}}^*(r)$.
\end{lemma}
In other words, Lemma~\ref{lem:DoQF_DF} states that the DMT achieved by the
DoQF protocol cannot be worse than the DMT achieved by the DF. The proof of 
Lemma \ref{lem:DoQF_DF} is given in Appendix \ref{anx:D1}.
 
\begin{lemma}\label{lem:DoQF_t0}
For any $r\in[0,1]$, the following inequalities hold true:
$\max\{0.5,r\}\leq t_{0,{\mathrm{DoQF}}}^*(r)\leq
t_{0,{\mathrm{DF}}}^*(r)$.
\end{lemma}
Here, $t_{0,{\mathrm{DF}}}^*(r)$ is the value of $t_0$ defined
by~\eqref{eq:DF_t0} which allows to achieve the DMT of the DF protocol.
The proof of Lemma \ref{lem:DoQF_t0} is given in Appendix  \ref{anx:D2}.

\begin{lemma}\label{lem:DoQF_delta}
Assume that $r>0.25$. The following holds true:
$0<\delta_{\mathrm{DoQF}}^*(r)<1-\left(1-
\frac{r}{t_{0,{\mathrm{DoQF}}}^*(r)}
\right)^+$. 
\end{lemma}
The proof of Lemma \ref{lem:DoQF_delta} is given in Appendix  \ref{anx:D3}.

These three lemmas will considerably simplify the derivation of
$d_{\mathrm{DoQF}}^*(r)$. Indeed, with the help of 
Lemma~\ref{lem:DoQF_t0} and Lemma~\ref{lem:DoQF_delta}, we will derive the
DMT of the DoQF firstly in the case when
$0.25<r\leq\frac{2(\sqrt{5}-1)}{9-\sqrt{5}}$, and secondly in the case when
$\frac{2(\sqrt{5}-1)}{9-\sqrt{5}}<r\leq 1$.
\begin{itemize}
\item $0.25<r\leq\frac{2(\sqrt{5}-1)}{9-\sqrt{5}}$.\\
We begin with the simplification of the DMT terms
$d_1\left(t_{0,{\textrm{DoQF}}}^*(r),r\right)$,
$d_2\left(t_{0,{\textrm{DoQF}}}^*(r),\delta_{\textrm{DoQF}}^*(r),r\right)$,
$d_3\left(t_{0,{\textrm{DoQF}}}^*(r),\delta_{\textrm{DoQF}}^*(r),r\right)$
and
$d_4\left(t_{0,{\textrm{DoQF}}}^*(r),\delta_{\textrm{DoQF}}^*(r),r\right)$.
The final DMT $d_{\textrm{DoQF}}^*(r)$ can then be deduced as the minimum of
the above terms.
Consider first the derivation of $d_1\left(t_{0,{\textrm{DoQF}}}^*(r),r\right)$.
Since Lemma~\ref{lem:DoQF_t0} states that
$t_{0,{\textrm{DoQF}}}^*(r)\leq t_{0,{\textrm{DF}}}^*(r)=\frac{2}{\sqrt{5}+1}$,
it follows from~\eqref{eq:DF_d1} that
\begin{equation}\label{eq:DoQF_d1_simp1}
 d_1\left(t_{0,{\textrm{DoQF}}}^*(r),r\right)=2-
\frac{r}{1-t_{0,{\textrm{DoQF}}}^*(r)}\:.
\end{equation}
We now proceed to the simplification of the expression of 
$d_2\left(t_{0,{\textrm{DoQF}}}^*(r),\delta_{\textrm{DoQF}}^*(r),r\right)$.
Thanks to Lemma~\ref{lem:DoQF_t0} and Lemma~\ref{lem:DoQF_delta},
we will prove that
\begin{equation}\label{eq:DoQF_d3_simp2}
d_2\left(t_{0,{\textrm{DoQF}}}^*(r),\delta_{\textrm{DoQF}}^*(r),
r\right)=(1-r)^++\max\left\{1-\frac{r}{t_{0,{\textrm{DoQF}}}^*(r)},1-r-
\delta_{\textrm{DoQF}}^*(r)\right\}\:.
\end{equation}
For that sake, refer to~\eqref{eq:DoQF_d3} and note that
proving~\eqref{eq:DoQF_d3_simp2} is equivalent to proving that
\begin{equation}\label{eq:t0_delta0_simp2}
\frac{r}{1-t_{0,{\textrm{DoQF}}}^*(r)}-\left(1-\frac{r}{t_{0,{\textrm{DoQF}}}
^*(r)}\right)^+-\frac{t_{0,{\textrm{DoQF}}}^*(r)}{1-t_{0,{\textrm{DoQF}}}^*(r)}
\delta_{\textrm{DoQF}}^*(r)\leq 1-r\:.
\end{equation}
In order to show that~\eqref{eq:t0_delta0_simp2} holds, we suppose to the
contrary that
$\frac{r}{1-t_{0,{\textrm{DoQF}}}^*(r)}-\left(1-\frac{r}{t_{0,{\textrm{DoQF}}}
^*(r)}\right)^+-\frac{t_{0,{\textrm{DoQF}}}^*(r)}{1-t_{0,{\textrm{DoQF}}}^*(r)}
\delta_{\textrm{DoQF}}^*(r)> 1-r$. Since
$\delta_{\textrm{DoQF}}^*(r)>0$ according to Lemma~\ref{lem:DoQF_delta},
the latter assumption leads to
\begin{equation}\label{eq:contradict1}
r>\frac{2t_{0,{\textrm{DoQF}}}^*(r)\left(1-t_{0,{\textrm{DoQF}}}
^*(r)\right)}{1+t_{0,{\textrm{DoQF}}}^*(r)
\left(1-t_{0,{\textrm{DoQF}}}^*(r)\right)}\:.
\end{equation}
Moreover, it is straightforward to show that
\begin{equation}\label{eq:contradict2}
\min_{0.5\leq t \leq \frac{2}{\sqrt{5}+1}}
\frac{2t\left(1-t\right)}{1+t\left(1-t\right)}>
\frac{2(\sqrt{5}-1)}{9-\sqrt{5}}\:,
\end{equation}
where the restriction to 
$0.5\leq t\leq t_{0,{\textrm{DF}}}^*(r)=\frac{2}{\sqrt{5}+1}$ is due to
Lemma~\ref{lem:DoQF_t0}.
Now, we can combine~\eqref{eq:contradict1} and~\eqref{eq:contradict2} in order
to get $r>\frac{2(\sqrt{5}-1)}{9-\sqrt{5}}\:,$
which contradicts the fact that $r\leq \frac{2(\sqrt{5}-1)}{9-\sqrt{5}}$. We
conclude that expression~\eqref{eq:DoQF_d3_simp2} holds true.

We can further simplify the expression~\eqref{eq:DoQF_d3_simp2}
by proving that $1-r-\delta_{\textrm{DoQF}}^*(r)\geq 
1-\frac{r}{t_{0,{\textrm{DoQF}}}^*(r)}$.
The proof of this point uses the same arguments as above and is thus omitted.
The term
$d_2\left(t_{0,{\textrm{DoQF}}}^*(r),\delta_{\textrm{DoQF}}^*(r),r\right)$
can finally be written as
\begin{equation}\label{eq:DoQF_d3_simp3}
d_2\left(t_{0,{\textrm{DoQF}}}^*(r),\delta_{\textrm{DoQF}}^*(r),
r\right)=2(1-r)^+-\delta_{\textrm{DoQF}}^*(r)\:.
\end{equation}
As for 
$d_3\left(t_{0,{\textrm{DoQF}}}^*(r),\delta_{\textrm{DoQF}}^*(r),r\right)$
given by~\eqref{eq:DoQF_d4}, it simplifies to
\begin{equation}\label{eq:DoQF_d4_simp}
d_3\left(t_{0,{\textrm{DoQF}}}^*(r),\delta_{\textrm{DoQF}}^*(r),r\right)=
4+\frac{t_{0,{\textrm{DoQF}}}^*(r)}{1-t_{0,{\textrm{DoQF}}}^*(r)}
\delta_{\textrm{DoQF}}^*(r)-\left(4+\frac{t_{0,{\textrm{DoQF}}}^*(r)}{1-t_{0
,{\textrm{DoQF}}}^*(r)} \right)\frac{r}{t_{0,{\textrm{ DoQF}}}^*(r)}
\end{equation}
The remaining task is to simplify the expression~\eqref{eq:DoQF_d5} which
defines
$d_4\left(t_{0,{\textrm{DoQF}}}^*(r),\delta_{\textrm{DoQF}}^*(r),r\right)$.
For that sake, we can resort to Lemma~\ref{lem:DoQF_DF} to prove
that 
$$d_4\left(t_{0,{\textrm{DoQF}}}^*(r),\delta_{\textrm{DoQF}}^*(r),r\right)=
(1-r)^++(1-\delta_{\textrm{DoQF}}^*(r))\:.$$ It follows that
$d_4\left(t_{0,{\textrm{DoQF}}}^*(r),\delta_{\textrm{DoQF}}^*(r),r\right)
\geq
d_2\left(t_{0,{\textrm{DoQF}}}^*(r),\delta_{\textrm{DoQF}}^*(r), r\right)$
and that it can thus be dropped from the derivation of the final DMT of the
DoQF. Now that the DMT terms
$d_1\left(t_{0,{\textrm{DoQF}}}^*(r),r\right)$,
$d_2\left(t_{0,{\textrm{DoQF}}}^*(r),\delta_{\textrm{DoQF}}^*(r),r\right)$
and
$d_3\left(t_{0,{\textrm{DoQF}}}^*(r),\delta_{\textrm{DoQF}}^*(r),r\right)$
have been expressed as functions of $t_{0,{\textrm{DoQF}}}^*(r)$ and
$t_{0,{\textrm{DoQF}}}^*(r)$, we can proceed to the determination of
$t_{0,{\textrm{DoQF}}}^*(r)$, $\delta_{\textrm{DoQF}}^*(r)$, and consequently
$d_{\textrm{DoQF}}^*(r)$.

\begin{itemize}
\item Determination of $\delta_{\textrm{DoQF}}^*(r)$:

Assume that $t_{0,{\textrm{DoQF}}}^*(r)$ has been already determined. 
It is straightforward to verify that $d_2\left(t,\delta,r\right)$
given by~\eqref{eq:DoQF_d3_simp3} is decreasing w.r.t $\delta$,
and that $d_3\left(t,\delta,r\right)$ given by~\eqref{eq:DoQF_d4_simp} is
increasing w.r.t $\delta$ on $\mathbb{R}^+$. Furthermore, 
$d_2\left(t,0,r\right)>d_3\left(t,0,r\right)$.
We conclude that 
$$
d_2\left(t_{0,{\textrm{DoQF}}}^*(r),\delta_{\textrm{DoQF}}^*(r),r\right)=
d_3\left(t_{0,{\textrm{DoQF}}}^*(r),\delta_{\textrm{DoQF}}^*(r),r\right)\:.
$$
Therefore, $\delta_{\textrm{DoQF}}^*(r)$ can be given as a function of
$t_{0,{\textrm{DoQF}}}^*(r)$ as follows
\begin{align}
\delta_{\textrm{DoQF}}^*(r)=\left(4-3t_{0,{\textrm{DoQF}}}^*(r)\right)
\frac{r}{t_{0,{\textrm{DoQF}}}^*(r)}-(2+2r)\left(1-t_{0,{\textrm{DoQF}}}
^*(r)\right)\:,\label{eq:DoQF_delta_simp}
\end{align}
which leads to
\begin{align}
&d_2\left(t_{0,{\textrm{DoQF}}}^*(r),\delta_{\textrm{DoQF}}^*(r),r\right)=
d_3\left(t_{0,{\textrm{DoQF}}}^*(r),\delta_{\textrm{DoQF}}^*(r),
r\right)=\nonumber\\
&2-2r+(2+2r)\left(1-t_{0,{\textrm{DoQF}}}^*(r)\right)
-\left(4-3t_{0,{\textrm{DoQF}}}^*(r)\right)
\frac{r}{t_{0,{\textrm{DoQF}}}^*(r)}\:.\label{eq:DoQF_d2_d3_simp}
\end{align}

\item Determination of $t_{0,\textrm{DoQF}}^*(r)$:

We can show in the same way that $t_{0,{\textrm{DoQF}}}^*(r)$ can be obtained
by writing 
\begin{equation}\label{eq:DoQF_t_simp}
d_1\left(t_{0,{\textrm{DoQF}}}^*(r),r\right)=
d_2\left(t_{0,{\textrm{DoQF}}}^*(r),\delta_{\textrm{DoQF}}^*(r),r\right)\:.
\end{equation}
Plugging the expression of $\delta_{\textrm{DoQF}}^*(r)$ 
from~\eqref{eq:DoQF_delta_simp} and the expression of
$d_2\Big(t_{0,{\textrm{DoQF}}}^*(r)$, $\delta_{\textrm{DoQF}}^*(r)$, $r\Big)$
from~\eqref{eq:DoQF_d2_d3_simp} into~\eqref{eq:DoQF_t_simp} leads
to equation~\eqref{eq:equation_t0} given in Theorem~\ref{the:DoQF_DMT} as
$$
2(1+r)t_{0,{\textrm{DoQF}}}^*(r)^3-
(4+5r)t_{0,{\textrm{DoQF}}}^*(r)^2+2(1+4r)t_{0,{\textrm{DoQF}}}^*(r)-4r=0\:.
$$
It can be shown after some algebra that the above equation admits a unique
solution $v^*(r)$ on $\left[0.5,\frac{2}{\sqrt{5}+1}\right]$ provided that
$r\leq \frac{2(\sqrt{5}-1)}{9-\sqrt{5}}$. This explains why the distinction
$r\leq\frac{2(\sqrt{5}-1)}{9-\sqrt{5}}$ and
$r>\frac{2(\sqrt{5}-1)}{9-\sqrt{5}}$ appears in Theorem~\ref{the:DoQF_DMT}.
Once the solution $v^*(r)$ to the above equation has been computed, 
then $d_{\textrm{DoQF}}^*(r)$, $t_{0,{\textrm{DoQF}}}^*(r)$ and
$\delta_{\textrm{DoQF}}^*(r)$ given respectively by~\eqref{eq:DoQF_dr},
\eqref{eq:DoQF_t0} and~\eqref{eq:DoQF_delta} can be easily obtained.
\end{itemize}

\item $\frac{2(\sqrt{5}-1)}{9-\sqrt{5}}<r\leq 1$.\\
In this case, we need to prove that
$d_{\textrm{DoQF}}^*(r)=d_{\textrm{DF}}^*(r)$.
To that end, we can show that $d_{\textrm{DoQF}}^*(r)>d_{\textrm{DF}}^*(r)$
leads to a contradiction. The proof of this point is based on
Lemmas~\ref{lem:DoQF_DF}, \ref{lem:DoQF_t0} and~\ref{lem:DoQF_delta} and is
omitted due to lack of space.
\end{itemize}
The proof of Theorem~\ref{the:DoQF_DMT} is thus completed.

\section{Proofs of Lemmas \ref{lem:DoQF_DF}, \ref{lem:DoQF_t0}, and \ref{lem:DoQF_delta}}

\subsection{Proof of Lemma~\ref{lem:DoQF_DF}}\label{anx:D1}

Assume that parameters $t_0$ and $\delta$ of the DoQF protocol are fixed such
that $t_0=t_{0,\textrm{DF}}^*(r)$ and
$\delta=1-\left(1-\frac{r}{t_{\textrm{DF}}^*(r)}\right)^+=
\frac{r}{t_{\textrm{DF}}^*(r)}$, where $t_{0,\textrm{DF}}^*(r)$ is defined
by~\eqref{eq:DF_t0}. In this case, equations~\eqref{eq:DF_d1},
\eqref{eq:DoQF_d3}, \eqref{eq:DoQF_d4} and~\eqref{eq:DoQF_d5} lead to
$d_1(t_0,r)=d_4(t_0,\delta,r)=d_{\textrm{DF}}^*(r)$ and
$d_2(t_0,\delta,r)=d_3(t_0,\delta,r)=2(1-r)^+$, meaning that
$d(t_0,\delta,r)=d_{\textrm{DF}}^*(r)$.

We conclude that the DoQF can be reduced to have the performance of DF by
choosing $t_0=t_{0,\textrm{DF}}^*(r)$ and
$\delta=\frac{r}{t_{0,\textrm{DF}}^*(r)}$. The final DMT
$d_{\textrm{DoQF}}^*(r)$ of the DoQF is therefore necessarily greater or equal
to $d_{\textrm{DF}}^*(r)$. The proof of Lemma~\ref{lem:DoQF_DF} is thus
completed.

\subsection{Proof of Lemma~\ref{lem:DoQF_t0}}\label{anx:D2}

Proving Lemma~\ref{lem:DoQF_t0} requires proving that the following three
inequalities hold: $r\leq t_{0,\textrm{DoQF}}^*(r)$,
$t_{0,\textrm{DoQF}}^*(r)$ $\leq t_{0,\textrm{DF}}^*(r)$ and
$0.5\leq t_{0,\textrm{DoQF}}^*(r)$.
Let us begin with the proof of the inequality $r\leq t_{0,\textrm{DoQF}}^*(r)$.
Assume to the contrary that $r> t_{0,\textrm{DoQF}}^*(r)$. In this case, 
$d_3(t_{0,\textrm{DoQF}}^*(r),\delta_{\textrm{DoQF}}^*(r),r)=0$ due
to~\eqref{eq:DoQF_d4}. This implies that the DMT of the DoQF satisfies
$d(t_{0,\textrm{DoQF}}^*(r),\delta_{\textrm{DoQF}}^*(r),r)
=d_3(t_{0,\textrm{DoQF}}^*(r),\delta_{\textrm{DoQF}}^*(r),r)=0$,
which is in contradiction with Lemma~\ref{lem:DoQF_DF}. We conclude that 
$r\leq t_{0,\textrm{DoQF}}^*(r)$ holds true.

We now show that the inequality
$t_{0,\textrm{DoQF}}^*(r)\leq t_{0,\textrm{DF}}^*(r)$ also holds true. For that
sake, note that the DMT $d_{\textrm{DF}}^*(r)$ of DF given
by~\eqref{eq:DF_DMT} can be written as a function of
$t_{0,\textrm{DF}}^*(r)$ defined by~\eqref{eq:DF_t0}:
\begin{equation}\label{eq:temp_df_d1}
d_{\textrm{DF}}^*(r)=2-\frac{r}{1-t_{0,\textrm{DF}}^*(r)}=
d_1\left(t_{0,\textrm{DF}}^*(r),r\right)\:,
\end{equation}
where the second equality in~\eqref{eq:temp_df_d1} can be easily checked by
referring to~\eqref{eq:DF_d1}. On the other hand,
\begin{equation}\label{eq:doqf_d1}
d_1\left(t_{0,\textrm{DoQF}}^*(r),r\right)\geq d_{\textrm{DoQF}}^*(r)
\end{equation}
due to~\eqref{eq:dr_min}. Furthermore, Lemma~\ref{lem:DoQF_DF} states that
\begin{equation}\label{eq:doqf_df_temp}
d_{\textrm{DoQF}}^*(r)\geq d_{\textrm{DF}}^*(r)\:.
\end{equation}
Combining~\eqref{eq:temp_df_d1}, \eqref{eq:doqf_d1} and~\eqref{eq:doqf_df_temp}
leads to $d_1\left(t_{0,\textrm{DoQF}}^*(r),r\right)\geq
d_1\left(t_{0,\textrm{DF}}^*(r),r\right)$. Since 
$d_1(t_0,r)=2-\frac{r}{1-t_0}$, we conclude that 
$t_{0,\textrm{DoQF}}^*(r)\leq t_{0,\textrm{DF}}^*(r)$ holds.

In order to prove that inequality $t_{0,\textrm{DoQF}}^*(r)\geq0.5$ holds, we
will show that the best DMT that can be achieved with $t_0<0.5$ \emph{i.e.,}
$\max_{t_0<0.5}d(t_0,\delta,r)$, is less or equal to the DMT that can be
achieved by choosing $t_0\geq0.5$. 
It can be shown after some algebra that
$$
\forall u\geq 0.5, \forall v<0.5,\:\:\:\: d_2(v,\delta,r)\leq
d_2(u,\delta,r)\:,
$$
where $d_2(u,\delta,r)$ is given by~\eqref{eq:DoQF_d3} and $d_2(v,\delta,r)$ is
given by~\eqref{eq:DoQF_d3_bis}. Furthermore, it is straightforward to show that
functions $t\mapsto d_3(t,\delta,r)$ and $t\mapsto d_4(t,\delta,r)$ defined
respectively by~\eqref{eq:DoQF_d4} and~\eqref{eq:DoQF_d5} are increasing w.r.t
$t$.
Finally, since $d_1(v,r)=2(1-r)^+$ for any $v<0.5$ due to~\eqref{eq:DF_d1}, then
$d(v,\delta,r)=\min\{d_2(v,\delta,r),d_3(v,\delta,r),d_4(v,\delta,r)\}$.
Putting all pieces together, we conclude that
$$
\forall u\geq 0.5, \forall v<0.5,\:\:\:\: d(v,\delta,r)\leq d(u,\delta,r)\:,
$$
which in turn means that $t_{0,\textrm{DoQF}}^*\geq 0.5$.

\subsection{Proof of Lemma~\ref{lem:DoQF_delta}}\label{anx:D3}

Lemma~\ref{lem:DoQF_delta} states that the
following two inequalities hold true for $r>0.25$:
\begin{center}
$\delta_{\mathrm{DoQF}}^*(r)<1-\left(1-
\frac{r}{t_{0,{\textrm{DoQF}}}^*(r)}\right)^+$ and
$0<\delta_{\mathrm{DoQF}}^*(r)$.
\end{center} 
Recall from our discussion in Appendix~\ref{app:der_d3} that the first
inequality is a necessary condition for the DMT of the DoQF protocol to be
greater or equal to the DMT of DF. We thus only need to prove the second
inequality. To that end, we will resort to Lemma~\ref{lem:DoQF_DF} which implies
that 
\begin{equation}\label{eq:ineq3}
d_3\left(t_{0,{\textrm{DoQF}}}^*(r),\delta_{\textrm{DoQF}}^*(r),
r\right)\geq d_{\textrm{DF}}^*(r)\:,
\end{equation}
where 
$d_3\left(t_{0,{\textrm{DoQF}}}^*(r),\delta_{\textrm{DoQF}}^*(r),r\right)=
4+\frac{t_{0,{\textrm{DoQF}}}^*(r)}{1-t_{0,{\textrm{DoQF}}}^*(r)}
\delta_{\textrm{DoQF}}^*(r)-\left(4+\frac{t_{0,{\textrm{DoQF}}}^*(r)}{1-t_{0,{
\textrm{DoQF}}}^*(r)} \right)\frac{r}{t_{0,{\textrm{ DoQF}}}^*(r)}$ due
to~\eqref{eq:DoQF_d4_simp}.
Consider first the case $\frac{\sqrt{5}-1}{\sqrt{5}+1}<r\leq1$. In this
case, $d_{\textrm{DF}}^*(r)=(1-r)(2-r)$ due to \cite{DMG}. 
Inequality~\eqref{eq:ineq3} is therefore equivalent to
$$
4+\frac{t_{0,{\textrm{DoQF}}}^*(r)}{1-t_{0,{\textrm{DoQF}}}^*(r)}
\delta_{\textrm{DoQF}}^*(r)-\left(4+\frac{t_{0,{\textrm{DoQF}}}^*(r)}{1-t_{0,{
\textrm{DoQF}}}^*(r)} \right)\frac{r}{t_{0,{\textrm{ DoQF}}}^*(r)}\geq
(1-r)(2-r)\:.
$$
It is straightforward to show that the above inequality is equivalent to
\begin{equation}\label{eq:ineq4}
\frac{t_{0,{\textrm{DoQF}}}^*(r)}{1-t_{0,{\textrm{DoQF}}}^*(r)}
\delta_{\textrm{DoQF}}^*(r)\geq
r^2+\left(\frac{4}{t_{0,{\textrm{DoQF}}}^*(r)}+
\frac{1}{1-t_{0,{\textrm{DoQF}}}^*(r)}-3\right)r-2\:.
\end{equation}
One can check after some algebra that the rhs of~\eqref{eq:ineq4} is
strictly positive for $\frac{\sqrt{5}-1}{\sqrt{5}+1}<r\leq1$. We conclude
that $\delta_{\textrm{DoQF}}^*(r)>0$ on this interval. The proof of the strict
positivity of
$\delta_{\textrm{DoQF}}^*(r)$ for $0.25<r\leq\frac{\sqrt{5}-1}{\sqrt{5}+1}$
can be done without difficulty in the same way, completing the proof of
Lemma~\ref{lem:DoQF_delta}.

\end{document}